\documentclass{aa}

\newcommand{\kms}{km s$^{-1}$}

\usepackage{natbib}

\usepackage{graphicx}

\usepackage{rotating}

\usepackage[varg]{txfonts}

\begin{document}


\title{Dust and gas in luminous proto-cluster galaxies at z=4.05:  the case for different cosmic dust evolution in normal and starburst galaxies}


\author{Q. Tan\inst{1,2,3}
	\and E. Daddi\inst{1}
	\and G. Magdis\inst{4}
	\and M. Pannella\inst{1}
	\and M. Sargent\inst{5}
	\and D. Riechers\inst{6}
	\and M. B\'{e}thermin\inst{1}
	\and F. Bournaud\inst{1}
	\and C. Carilli\inst{7}
	\and E. da Cunha\inst{8}
	\and H. Dannerbauer\inst{9}
	\and M. Dickinson\inst{10}
	\and D. Elbaz\inst{1}
	\and Y. Gao\inst{2}
	\and J. Hodge\inst{8}
	\and F. Owen\inst{7}
	\and F. Walter\inst{8}
}
\institute{CEA Saclay, DSM/Irfu/Service d'Astrophysique, Orme des Merisiers, F-91191 Gif-sur-Yvette Cedex, France
	\and Purple Mountain Observatory \& Key Laboratory for Radio Astronomy, Chinese Academy of Sciences, Nanjing 210008, China
	\and Graduate University of the Chinese Academy of Sciences, 19A Yuquan Road, Shijingshan District, Beijing 10049, China
	\and Department of Physics, University of Oxford, Keble Road, Oxford OX1 3RH, UK
	\and Astronomy Center, Dept. of Physics \& Astronomy, University of SUssex, Brighton BN1 9QH, UK
	\and Department of Astronomy, Cornell University, 220 Space Sciences Building, Ithaca, NY 14853, USA
	\and National Radio Astronomy Observatory, P.O. Box O, Socorro, NM 87801, USA
         \and Max-Planck Institute for Astronomy, K\"{o}nigstuhl 17, 69117 Heidelberg, Germany
	\and Universit\"{a}t Wien, Institut f\"{u}r Astrophysik, T\"{u}rkenschanzstrasse 17, 1180 Vienna, Austria
	\and National Optical Astronomical Observatory, 950 North Cherry Avenue, Tucson, AZ 85719, USA
	}




\date{Received 30 March 2014/Accepted 11 July 2014}

\abstract{We measure the dust and gas content of the three sub-millimeter galaxies (SMGs) in the GN20 proto-cluster at $z=4.05$ using new IRAM Plateau de Bure interferometer (PdBI) CO(4-3) and 1.2--3.3 mm continuum observations. All these three SMGs are heavily dust obscured, with UV-based star formation rate (SFR) estimates significantly smaller than the ones derived from the bolometric infrared (IR), consistent with the spatial offsets revealed by HST and CO imaging. 
Based also on evaluations of the specific SFR, CO-to-H$_2$ conversion factor and gas depletion timescale, we classify all the three galaxies as starbursts (SBs), although with a lower confidence for GN20.2b that might be a later stage merging event. We place our measurements in the context of the evolutionary properties of main sequence (MS) and SB galaxies. ULIRGs have 3--5 times larger
$L'_{\rm CO}/M_{\rm dust}$ and $M_{\rm dust}/M_\star$ ratios than $z=0$ MS galaxies, but by $z\sim2$ the difference appears to be blurred, probably due to differential metallicity evolution. SB galaxies
appear to  slowly evolve in their $L'_{\rm CO}/M_{\rm dust}$ and $M_{\rm dust}/M_\star$ ratios all the way to $z>6$ (consistent with rapid enrichment of SB events), while MS galaxies rapidly increase in
$M_{\rm dust}/M_\star$  from $z=0$ to 2 (due to gas fraction increase, compensated by a decrease of metallicities).  While no IR/submm continuum detection is available for indisputably normal massive galaxies at $z>2.5$, we show that if metallicity indeed decrease rapidly for these systems  at $z>3$ as claimed in the literature, we should expect a strong decrease of their $M_{\rm dust}/M_\star$, consistent with recent PdBI and ALMA upper limits. 
We conclude that the $M_{\rm dust}/M_\star$ ratio could  be a powerful tool for distinguishing starbursts from normal galaxies at $z>4$.
} 

\titlerunning{Dust and gas in luminous proto-cluster galaxies at z=4.05}
\authorrunning{Tan et al.}

\maketitle

\section{Introduction}

Submillimeter (submm) and millimeter observations are efficient in detecting and studying dusty, star-forming galaxies, due to the effect of negative $K$-correction, which results in nearly constant observed brightness for galaxies with same infrared (IR) luminosity over a broad range of redshifts. However, most current deep submm surveys are limited to the brightest sources and submm-selected galaxies \citep[SMGs;][]{blain02}, due to the limited spatial resolution and sensitivity of submm observations. SMGs are massive, highly dust obscured galaxies with extreme star formation rates (SFRs) of order 10$^3$ $M_\odot$ yr$^{-1}$ \citep[e.g., review by][]{blain02,casey14}, and are generally thought to represent the progenitors of local massive elliptical galaxies. While spectroscopic studies of SMGs originally gave a median redshift of {\em z}$\sim$2.5 \citep{chapman05}, recent deep submm/mm continuum and radio observations show evidence for a significant population of higher redshift massive starbursts \citep[SBs; e.g.,][]{dannerbauer04,smolcic12,swinbank13,dowell14}, extending the redshift peak beyond $z=3$. A substantial number of $z>4$ SMGs have been identified  to date \citep[e.g.,][]{dannerbauer08,daddi09a,daddi09b,capak08,capak11,schinnerer08,coppin09,knudsen10,riechers10,riechers13,smolcic11,walter12,combes12,vieira13}. The surface density of these galaxies is found to be significantly higher than that expected from theoretical models \citep[e.g.,][]{baugh05,hayward13}, suggesting that current models of galaxy formation underpredict the number of high-redshift starbursts.  

Observations of the molecular gas in high-redshift galaxies reveal that while SMGs are highly gas-rich systems \citep{tacconi08}, the gas fractions of these systems are comparable to those of typical massive galaxies at similar epochs \citep[$\sim$40-60\%;][]{daddi08,daddi10}, implying that SMGs have higher star formation efficiencies \citep[SFEs;][]{daddi10,genzel10}. However, these results are complicated by the large uncertainties associated with the CO-to-H$_2$ conversion factor $\alpha_{\rm CO}$ \footnote{{\bf $\alpha_{\rm CO} = M_{\rm H_2} / L'_{\rm CO}$}, with units of M$_\odot$ (K km s$^{-1}$ pc$^2$)$^{-1}$, which are omitted from the text for brevity. Note that the contribution from Helium is included in the $\alpha_{\rm CO}$ estimates.}, which likely changes between normal disk galaxies and starbursts \citep[see review by][]{bolatto13,carilli13}. In the literature, an ``ULIRG-like'' value of $\alpha_{\rm CO}=0.8$ (Downes \& Solomon 1998; but see \citealt{papadopoulos12} for a higher value of $\alpha_{\rm CO}$) is widely adopted for SMGs due to the lack of direct measurements, while a value of $\alpha_{\rm CO}\sim$ 4 is favored for Milky Way and normal galaxies. This carries a significant uncertainty since high redshift SMGs may be dramatically different from local ULIRGs, given the more extended gas distribution and different physical conditions revealed in some SMGs \citep{riechers11,ivison11,carilli13,scoville14}. Therefore, it is of significant importance to obtain a direct calibration of $\alpha_{\rm CO}$,  since well-determined molecular gas masses are critical to study the variations in physical properties across the galaxy populations at high redshift. 

Because of the extreme high specific star formation rates (sSFRs), some of the most luminous SMGs are placed as outliers above the main sequence (MS) of star formation, which is a tight correlation observed between the stellar mass and the SFR over a broad range of redshifts \citep[e.g.,][and references therein]{noeske07,elbaz07,daddi07b,rodighiero10}. While galaxies on the MS are thought to form stars gradually with a long duty cycle and represent the bulk of the galaxy population, starbursts exhibit very intense and rapid star formation activity, likely driven by mergers \citep[e.g.,][]{daddi07a,daddi07b,tacconi08,tacconi10,elbaz11,rodighiero11}. Recent studies on the molecular gas of $z > 3$ Lyman break galaxies (LBGs) found these galaxies to be rather deficient in CO emission for their star formation activity \citep{magdis12b,tan13}. Similar results have also been reported for a luminous LBG at $z=6.595$ called ``Himiko'', for which the 1.2 mm dust continuum and [CII] 158 $\mu$m emission are much lower than predicted by local correlations and measured SFRs \citep{ouchi13}. It has been found that normal galaxies at $z >$ 3 are increasingly metal poor, with metallicities dropping by about 0.6 dex as compared to local galaxies of similar stellar mass \citep{mannucci10,sommariva12,troncoso13}. This may suggest that metallicity effects could be a probable explanation for the deficit of CO emission, since the photodissociation of CO by far-UV radiation is enhanced at low metallicity \citep{leroy11,genzel12,narayanan12,bolatto13}. The decrease of CO emission in $z > 3$ normal galaxies for their IR luminosity is also predicted by simulations with a galaxy-formation model \citep{lagos12}, a result driven by the low metallicities in such objects.  Similar detailed study of a local metal poor star-forming galaxy, I Zw 18, concluded that it would be much harder than hitherto anticipated to detect gas and dust in high-redshift galaxies like Himiko \citep[several tens of days of integration with the complete ALMA; see][]{fisher14}, if assuming I Zw 18 is an analog of primitive galaxy population in the early Universe. 
Dust is expected and observed to be well mixed with gas in the interstellar medium (ISM), because it is composed of metals and regulates the gas phase abundances of the elements through accretion and destruction processes \citep{draine07}. The comparison of dust and gas properties of galaxies at different redshifts is thus crucial to explore the interplay between dust, gas, and metals in the ISM, and allow us to achieve a better understanding of galaxy evolution throughout cosmic time. For galaxies at $z>3$, to date only very few luminous SMGs have been detected in both dust continuum and gas emission \citep[e.g.,][]{dannerbauer08,daddi09a,daddi09b,coppin09,coppin10,walter12,riechers13}, while no dust continuum detection is available for indisputably normal galaxies at $z>3.5$. 

GN20 is one of the brightest SMGs in the GOODS-N field \citep{pope06}, of which the redshift ($z=4.055$) was established by a serendipitous detection of its CO(4-3) emission \citep{daddi09b}. Two additional SMGs, GN20.2a and GN20.2b, were found to lie within  $\sim$25$''$ of GN20 (projected physical separation $\sim$ 170 kpc) and have redshifts of $z\sim$ 4.055$\pm$0.005. These two galaxies are separated by only a few arcseconds and hence are not spatially separated from each other in existing submm images (e.g., SCUBA 850~$\mu$m). \citet{daddi09b} also found 14 B-band dropouts (roughly $z\sim4$) lying within 25$''$ from GN20, which corresponds to an overdensity of 5.8$\sigma$ in the GOODS-N field, suggesting a massive proto-cluster environment at $z\sim4.05$, just 1.6 Gyr after the Big Bang. All three massive SMGs in the GN20 proto-cluster have been detected in CO emission, indicative of large amounts of molecular gas \citep{daddi09b,carilli10,carilli11,hodge12,hodge13}, feeding vigorous ongoing star formation (SFR $\sim$ a few to ten times 100 $M_\odot$ yr$^{-1}$). The deep, high-resolution CO(2-1) observations reveal a clumpy, extended gas disk (14$\pm$4 kpc in diameter) for GN20 \citep{hodge12}, and extended gas reservoirs ($\sim$ 5--8 kpc) for GN20.2a and GN20.2b \citep{hodge13}.  \citet{hodge12,hodge13} have attempted to constrain the estimate of CO-to-H$_2$ conversion factor by dynamical analysis, deriving $\alpha_{\rm CO}$ of $\sim$ 1--2 for these three galaxies.  For the dust continuum emission, however, only GN20 has been reported in {\it Herschel} and (sub)mm detections \citep{daddi09b,magdis11}. Here we use PACS and PdBI mm data for GN20.2a and GN20.2b to study the dust properties of these two galaxies. By combining with the molecular gas properties and dynamical analysis, we aim to achieve a more comprehensive understanding of the nature of the massive SMGs in the GN20 proto-cluster environment. We further investigate the metallicity effects on molecular gas and dust emission by comparison of CO luminosity-to-dust mass ratio and dust-to-stellar mass ratio between normal galaxies and starbursts.

This paper is organized as follows. In Sect.~\ref{obs} we present the new PdBI CO(4-3) observations and the reduction of data for the SMGs GN20, GN20.2a, and GN20.2b. Section~\ref{results} presents the results of CO and millimeter continuum observations, the methods used to compute SFR, dust mass, stellar mass, metallicity, dynamical mass, and CO-to-H$_2$ conversion factor. This section also describes the derived physical properties including sSFR, SFE, radio-IR correlation constraints, gas fraction, and gas depletion timescales. In Sect.~\ref{nature} we discuss the nature of GN20, GN20.2a, and GN20.2b based on the physical properties of optical morphology, molecular gas, dust, and dynamical constraints. We further discuss the implications for the cosmic evolution of dust content in galaxies in sect.~\ref{implication}. Finally, we summarize our results in Sect.~\ref{summary}. We adopt a cosmology with $H_0$=71 km s$^{-1}$ Mpc$^{-1}$, $\Omega_{\rm M}$=0.3, $\Omega_\Lambda$=0.7, and a Chabrier (2003) initial mass function (IMF) throughout the paper.

\begin{figure*}[htbp]
\centering
\includegraphics[width=0.32\textwidth]{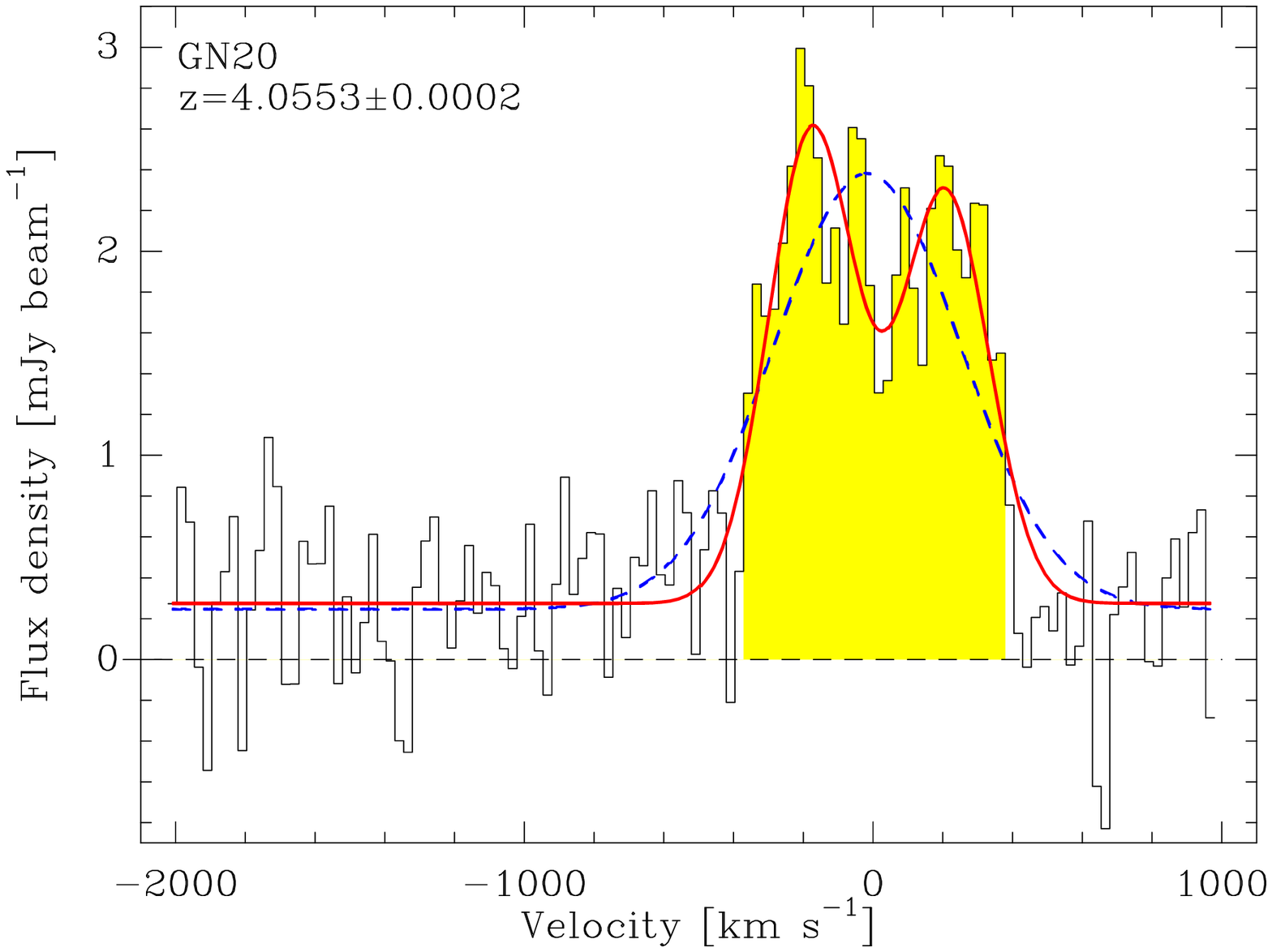}
\includegraphics[width=0.32\textwidth]{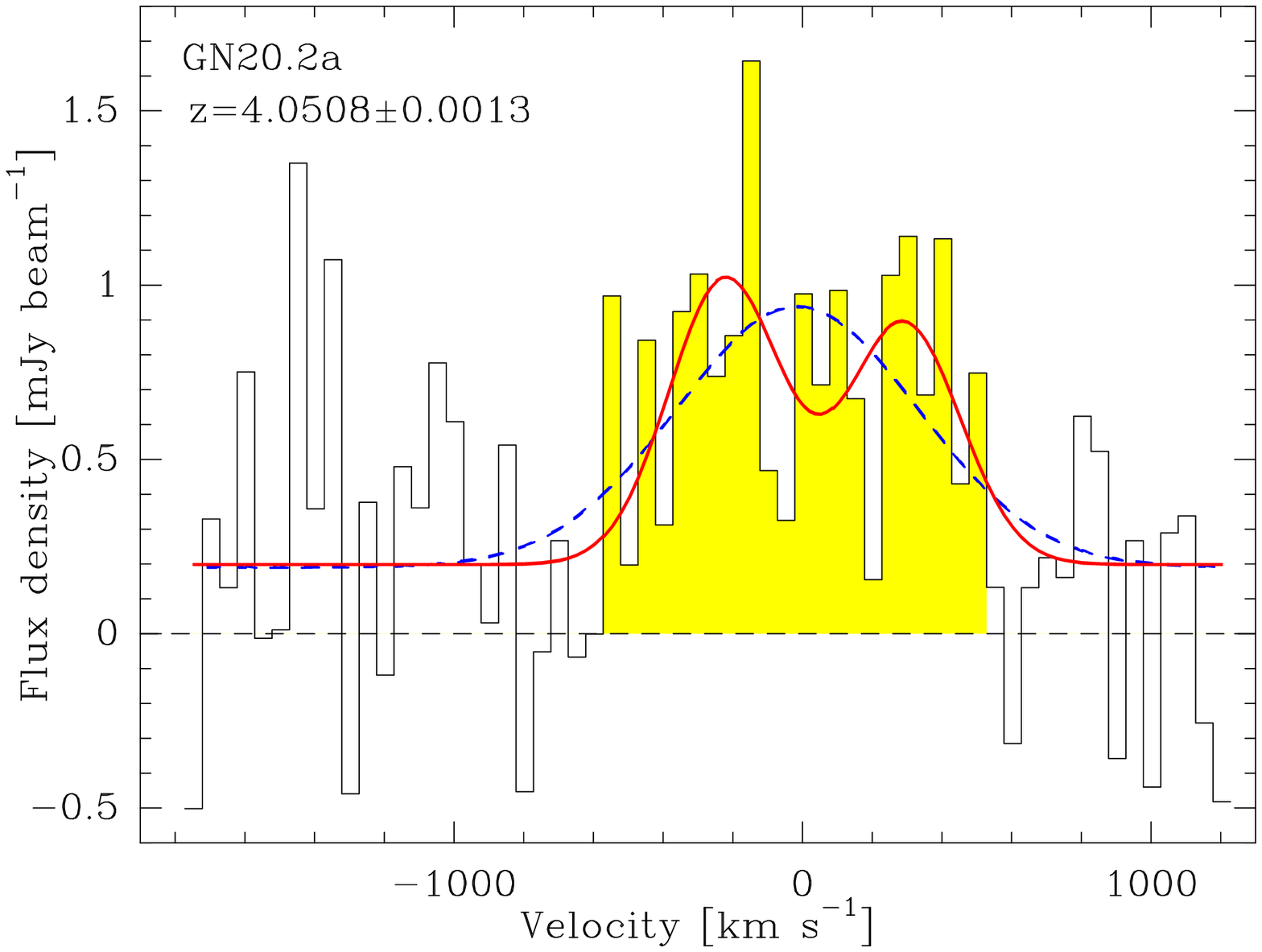}
\includegraphics[width=0.32\textwidth]{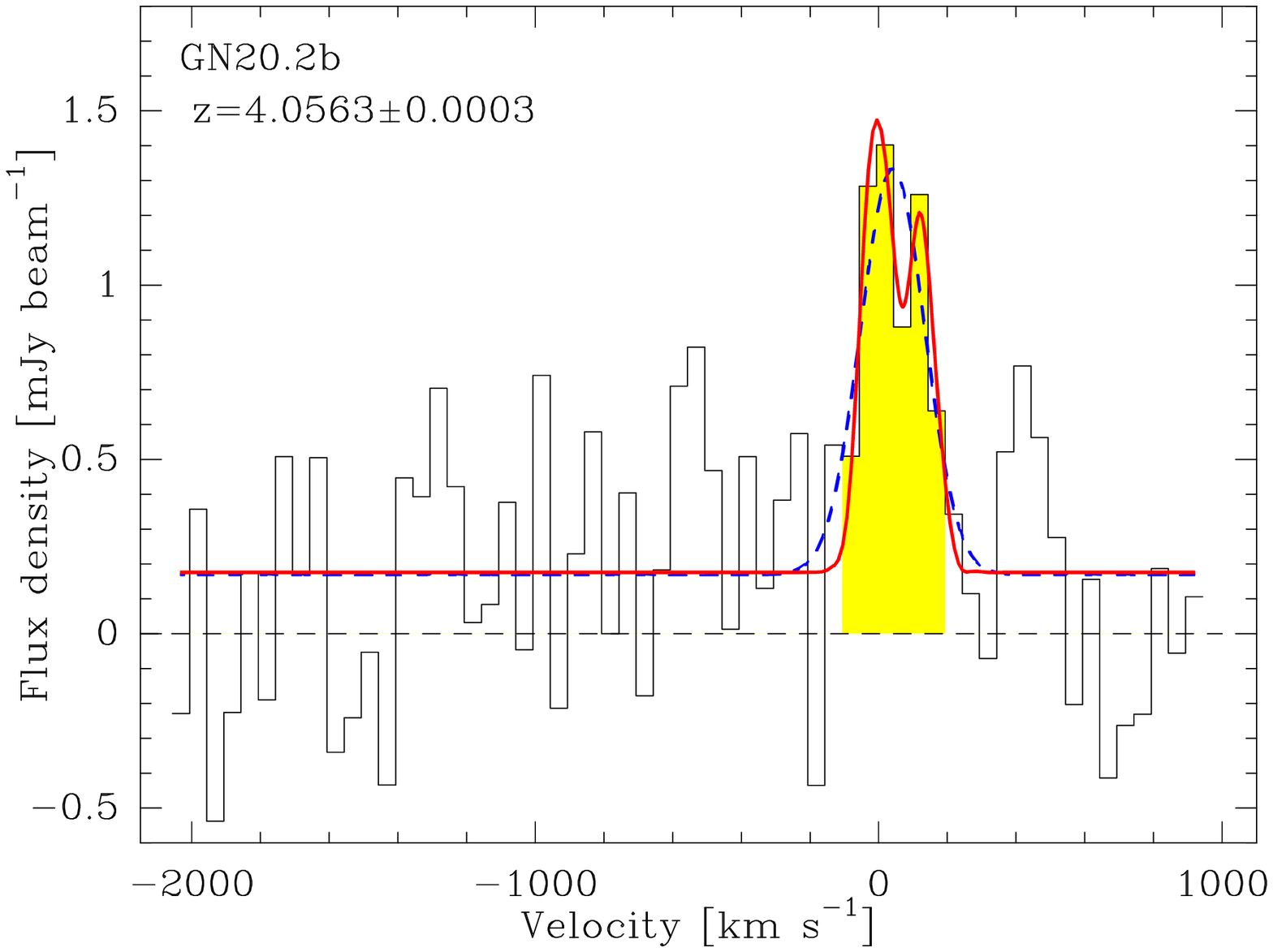}
\caption{CO(4-3) spectra binned in steps of 25 km s$^{-1}$ for GN20 (left), 50 km s$^{-1}$ for GN20.2a (middle) and GN20.2b (right). The yellow color indicates the velocity ranges where line emission is detected. These regions have been used to derive the integrated fluxes. The red lines show best-fitting double Gaussian profiles to the spectra, and the blue dashed lines show the fits with a single Gaussian.
Zero velocities correspond to the redshifts listed in Table~1. \label{fig1}}
\end{figure*}

\begin{table*}
\caption{Observed and derived CO emission properties}
\label{tbl1}
\centering
\small
{\renewcommand{\arraystretch}{1.5}
\begin{tabular}{lccccccc}
\hline\hline
Source & 
R.A.$_{\rm CO}$ & 
Decl.$_{\rm CO}$ & 
$z_{\rm CO}$ & 
$z_{\rm keck}$  & 
FWHM & 
$I_{\rm CO (4-3)}$ & 
$L'_{\rm CO (4-3)}$  \\
 & 
(J2000) & 
(J2000) & 
 & 
 & 
(km s$^{-1}$) & 
(Jy km s$^{-1}$) & 
(K km s$^{-1}$ pc$^2$) \\
\hline

GN20 & 12:37:11.91 & 62:22:12.1 & 4.0553$\pm$0.0002 & 4.06$\pm$0.02  & 583$\pm$36 & 1.68$\pm$0.10 & 6.56$\times$10$^{10}$  \\
GN20.2a & 12:37:08.76 & 62:22:01.6 & 4.0508$\pm$0.0013 & 4.059$\pm$0.007  & 760$\pm$180 & 0.65$\pm$0.08 & 2.70$\times$10$^{10}$   \\
GN20.2b & 12:37:09.68 & 62:22:02.2 & 4.0563$\pm$0.0003 & ...  & 220$\pm$43 & 0.27$\pm$0.04 & 1.06$\times$10$^{10}$  \\

\hline
\end{tabular}
}
\end{table*}

\section{Observations and Data Reductions}\label{obs}
We used the IRAM PdBI to observe the CO(4-3) emission in the GN20 field in the AB, C and D configurations. The AB configuration observations were pointed close to a nearby $z=1.5$ galaxy and have been reported in \citet{daddi09b}, while the new C and D configuration observations, carried out in June 2009 and January-April 2013, were centered near a $z=4.058$ LBG, which is located $\sim16''$ south of GN20 \citep{daddi09b}. Both GN20 and its companions, GN20.2a and GN20.2b, are within the primary beam of the PdBI  observations. The primary beam attenuation (PBA) for GN20, GN20.2a, and GN20.2b are 1.29, 1.44, and 1.20 in C configuration observations, and 1.27, 1.18, and 1.06 in D configuration observations, respectively. All observations were tuned at a central frequency of 91.375 GHz and performed in dual polarization mode with six antennas available. For the observations performed in 2013, a 3.6 GHz WIDEX correlator was used. Further details on the observations can be found in \citet{tan13}. 

We reduced the data with the GILDAS software packages CLIC and MAPPING. After flagging bad and high phase noise data, and correcting for the PBA, the total effective on-source integration time for GN20, GN20.2a, and GN20.2b are 14.6 h, 11.9 h, and 15.1 h, respectively. We combined the C and D configuration data with published AB configuration observations to maximize the sensitivity.  The spectra shown in Fig.~\ref{fig1} were extracted from the combined UV data by using a circular Gaussian model with a fixed full width at half maximum (FWHM) of 0.72$''$, 0.53$''$,  0.88$''$ for GN20, GN20.2a, and GN20.2b \citep{carilli10,hodge13}, respectively. The CO(4-3) images integrating over the velocity range where emission was detected and the 3.3 mm maps averaging over the line-free channels for the three galaxies are shown in Fig.~\ref{fig2}. In order to avoid the contamination from the side lobes of the bright GN20 galaxy, the emission at the position of GN20 was fitted in the UV data with a circular Gaussian source model and subtracted before creating the images of GN20.2a and GN20.2b shown in Figure~\ref{fig2}. The sensitivities of combined data set at the position of GN20, GN20.2a, and GN20.2b are 0.41,  0.63, and 0.52 mJy beam$^{-1}$ per 25 \kms \ channel. 

We make use of the 1.2 mm and 2.2 mm continuum observations of the GN20 field with the PdBI. The 2.2 mm continuum data of our three galaxies haven been reported in \citet{carilli10}. More details on the 1.2 mm observations and reduction will be given in  Riechers et al. (in prep.).

\section{Results and Analysis}\label{results}
\subsection{CO(4-3) emission properties}
The CO detections of GN20, GN20.2a, and GN20.2b at various transitions have been presented in some recent studies \citep{daddi09b,carilli10,carilli11,hodge12,hodge13}. Compared to these published work, our new deep CO(4-3) data allow us to improve the constraint on the CO redshift, flux density, and line width, which is a vital parameter for the determination of dynamic mass. In addition, we put useful constraints on the 3 mm continuum emission for these SMGs. Table~\ref{tbl1} summarizes the observed and derived CO properties for these three galaxies.

\begin{figure*}[htbp]
\centering
\includegraphics[width=0.55\textwidth]{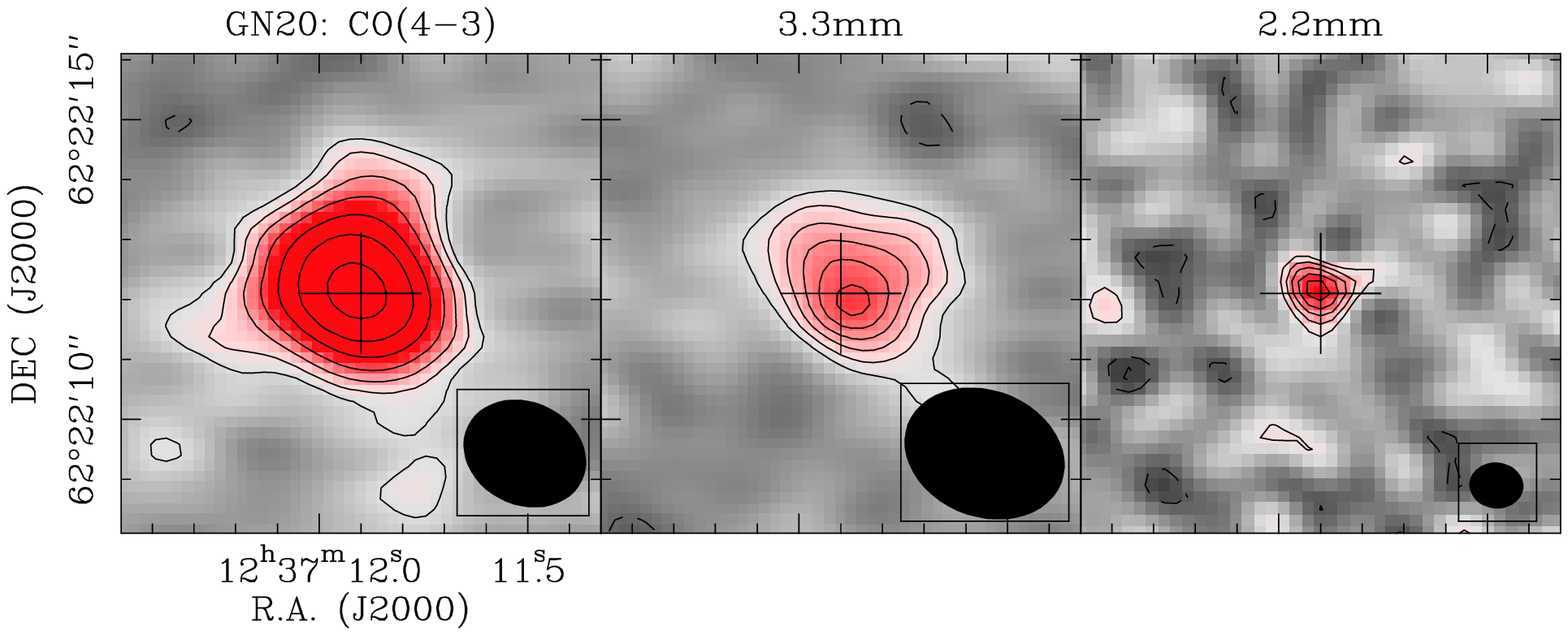}\\
\includegraphics[width=0.55\textwidth]{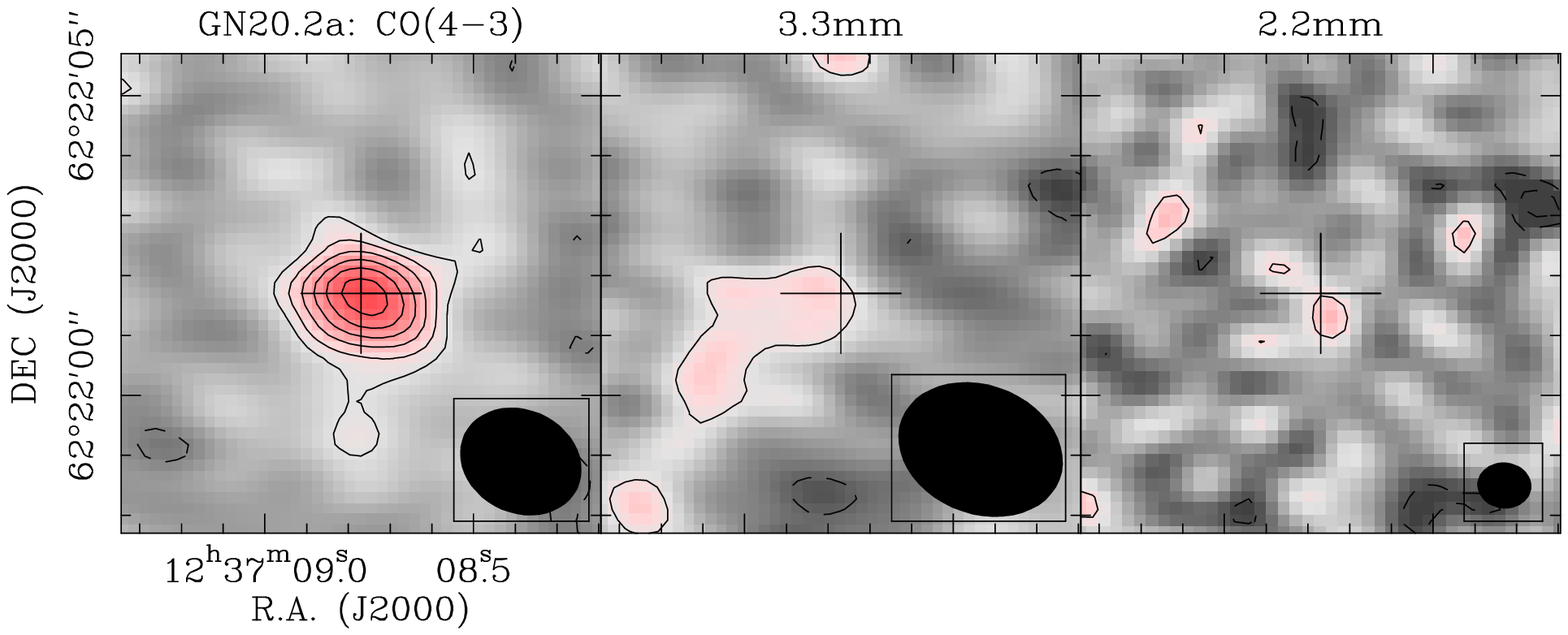}\\
\includegraphics[width=0.55\textwidth]{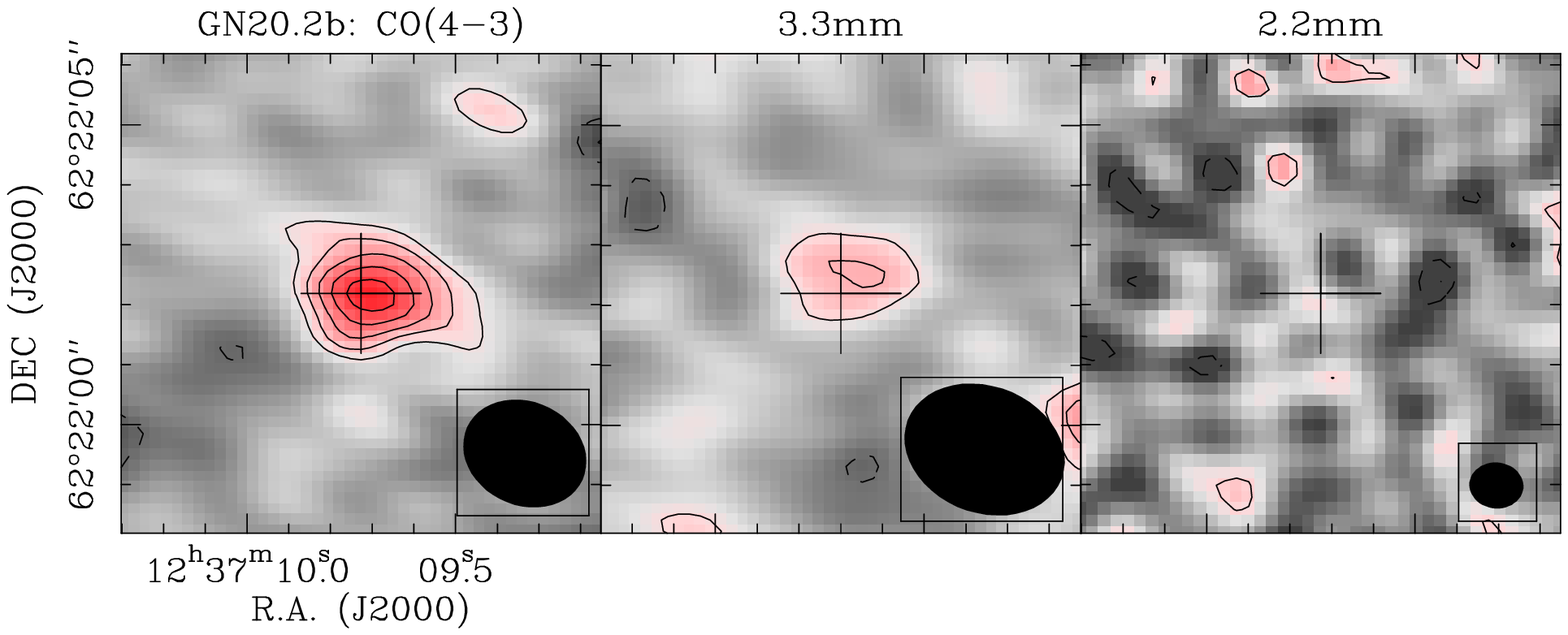}
\caption{PdBI maps (from left to right) of CO(4-3) line, 3.3 mm, and 2.2 mm continuum emission for GN20 (top), GN20.2a (middle), and GN20.2b (bottom). The CO(4-3) images are averaged over the observed velocity range of CO emission (see Fig.~\ref{fig1}) and were cleaned. The 2.2 mm and 3.3 mm continuum maps were created by averaging over the line-free emission channels and were not cleaned given the relatively low S/N. The crosses show the positions where we extract CO spectra. For GN20, the contours level of CO(4-3) are shown increase by a factor of 1.5 starting at $\pm2 \sigma$, with positive (negative) contours shown as solid (dashed) lines. For all the continuum maps and CO(4-3) maps of GN20.2a and GN20.2b, contours start at $\pm2 \sigma$ in steps of 1$\sigma$. Beam sizes are displayed in the lower right corner of each integrated map. \label{fig2}}
\end{figure*}

To measure the velocity centroid and line width of CO emission of GN20, we fitted Gaussians to the observed spectra (Fig.~\ref{fig1}), which is extracted at the fixed CO position from \citet{daddi09b}, allowing for the presence of faint underlying continuum. The CO(4-3) spectrum of GN20 shows clear evidence of double-peaked structure. A single Gaussian fit to the spectrum shown in Fig.~\ref{fig1} yields a peak flux density of 2.13$\pm$0.14 mJy and an FWHM of 630$\pm$51 \kms, consistent with \citet{daddi09b}. We also fitted the spectrum with a double Gaussian function by fixing the FWHM in each component to the same value. The peak flux densities for the two components are 2.33$\pm$0.17 mJy and 2.02$\pm$0.16 mJy, respectively. We find that the total FWHM (583$\pm$36 \kms) derived from a double Gaussian fitting is better constrained than that from a single one. The redshift derived from a double Gaussian is  4.0553$\pm$0.0002, consistent with the previous measurements \citep{daddi09b,carilli11,hodge12}. A circular Gaussian model fit with a fixed FWHM of 0.72$''$ (we here adopt the FWHM of the CO(6-5) image measured by Carilli et al. 2010) to the CO(4-3) image of GN20 (Fig.~\ref{fig2}) yields a velocity-integrated flux of $I_{\rm CO}=1.68\pm0.10$ Jy \kms, in line with the measurements in \citet{daddi09b}. 

For GN20.2a, the fits with a single Gaussian and two Gaussian functions yield FWHM of 820$\pm$237 \kms \ and 763$\pm$180 \kms, respectively. The spectrum is extracted at the fixed CO position from \citet{daddi09b}. We find the CO spectrum of GN20.2a appears to be better described by a double Gaussian profile (Fig.~\ref{fig1}), given the relatively low uncertainties of FWHM derived from double Gaussian fitting and the rapid decrease of the flux at the edges of the spectrum, though there is no clear indication of  double-peaked emission in the spectrum. The favored fit with two Gaussian functions could in principle suggests either the existence of a rotating component, or kinematically distinct components undergoing a merger within the system. We will further discuss the possibility in the following section. The two components in the double Gaussian fitting have peak flux densities of 0.82$\pm$0.21 mJy and 0.69$\pm$0.22 mJy, respectively. A Gaussian fit to the CO spectrum gives a redshift of 4.0508$\pm$0.0013. We used a circular Gaussian model with a fixed FWHM of 0.53$''$ (derived from CO(2-1) image in Hodge et al. 2013) to fit the CO(4-3) map and derive a velocity-integrated flux of 0.65$\pm$0.08 Jy \kms. All the measurements derived for GN20.2a are consistent with published results \citep{daddi09b,carilli11,hodge13}, but exhibit significantly improved sensitivity and signal-to-noise (S/N) ratio.

The CO(4-3) spectrum of GN20.2b has been presented in \citet{hodge13}. Compared to the large uncertainty in the line width fitted in  \citet{hodge13}, our deeper CO(4-3) data provide a much better constraint on the estimate of the line width. Similarly, we performed Gaussian fits to the spectrum (Fig.~\ref{fig1}) extracted at the fixed position of 1.2 mm emission in GN20.2b (Riechers et al., in prep.) with a single Gaussian and double Gaussian functions, respectively. Both Gaussian fits give a similar FWHM with value of $\sim220\pm40$ \kms, which is much smaller than GN20 and GN20.2a, and also the SMGs at $z\sim 2-4$ with mean FWHM of 470$\pm$80 \kms  \citep{bothwell13}.  The Gaussian fit to the spectrum has a peak flux density of 1.17$\pm$0.19 mJy and a redshift of 4.0563$\pm$0.0003. The velocity-integrated flux density derived from the  CO(4-3) map with a circular Gaussian model  (fixed FWHM of 0.88$''$; see Hodge et al. 2013) is 0.27$\pm$0.04 Jy \kms. 

We calculate the CO(4--3) line luminosities (in K km s$^{-1}$ pc$^2$) using the standard relation given by \citet{solomon97}:
\begin{equation}
L'_{\rm CO} = 3.25 \times 10^7 S_{\rm CO} \Delta V \nu_{\rm obs}^{-2} D^2_{\rm L} (1+z)^{-3}
\end{equation}
where $S_{\rm CO} \Delta V$ is the velocity-integrated line flux in Jy km s$^{-1}$, $\nu_{\rm obs}$ is the observed frequency in GHz and $D_{\rm L}$ is the luminosity distance in Mpc. With $L'_{\rm CO(2-1)}$ measured by \citet{carilli11}, we find CO(4-3)/CO(2-1) line brightness temperature ratios of $r_{43/21}$ = 0.41, 0.36, and 0.44 for GN20, GN20.2a, and GN20.2b, respectively. These are consistent with the mean ratio ($\sim$0.48$\pm$0.10) measured for SMGs at $z\sim 2-4$ \citep{bothwell13}.

\begin{table*}
\caption{Far-IR and (sub)mm properties of GN20, GN20.2a and GN20.2b} 
\label{tbl2}
\centering
\small\addtolength{\tabcolsep}{-3.5pt}

{\renewcommand{\arraystretch}{1.5}
\begin{tabular}{lccccccccccc}
\hline\hline

Source & 
$S_{\rm 100\mu m}$ & 
$S_{\rm 160\mu m}$  & 
$S_{\rm 250\mu m}\tablefootmark{a}$ & 
$S_{\rm 1.2 mm}$ & 
$S_{\rm 2.2 mm}$ & 
$S_{\rm 3.3 mm}$ & 
$S_{\rm 1.4 GHz}\tablefootmark{b}$ &
log$M_{\rm dust}$ & 
$\langle U \rangle$ & 

log$L_{\rm IR}$ & 
SFR$_{\rm IR}$ \\
 & 
(mJy) & 
(mJy) & 
(mJy) & 
(mJy)  & 
(mJy)  &
(mJy) &  
($\mu$Jy) &
 ($M_\odot$)  & 
& 

 ($L_\odot$) &
 ($M_{\odot}$yr$^{-1}$)\\

\hline
GN20$\tablefootmark{c}$ & 0.70$\pm$0.42 & 5.45$\pm$1.02 & 18.66$\pm$2.70 & 8.47$\pm$0.79 & 0.95$\pm$0.14 & 0.229$\pm$0.036 & 66.4$\pm$6.6 & 9.72$_{-0.03}^{+0.04}$ & 27.2$_{-2.2}^{+2.6}$ &  13.27$_{-0.02}^{+0.02}$ & 1860$\pm$90 \\

GN20.2a & 0.12$\pm$0.44 & 1.24$\pm$1.38 & ... & 3.83$\pm$0.48 & 0.52$\pm$0.23 & 0.177$\pm$0.071 & 138.5$\pm$3.5 & 9.39$_{-0.13}^{+0.16}$ & 26.4$_{-14.1}^{+8.1}$ &  12.90$_{-0.05}^{+0.02}$  & 800$\pm$70 \\

GN20.2b & 0.61$\pm$0.46 & 2.25$\pm$1.48 & ... & 3.25$\pm$0.52 & 0.25$\pm$0.16 & 0.114$\pm$0.035  & 14.2$\pm$4.1 & 9.32$_{-0.09}^{+0.17}$ & 25.8$_{-12.5}^{+6.3}$ &  12.84$_{-0.10}^{+0.04}$ & 690$\pm$100 \\ 

\hline
\end{tabular}
}
\tablefoot{
\tablefoottext{a}{The SPIRE 250$\mu$m detections for GN20.2a and GN20.2b are blended, due to the small separation ($\sim$6.8$''$) between these two galaxies.}
\tablefoottext{b}{ JVLA 1.4GHz flux density from new deep radio measurements (Owen et al., in prep.). { For GN20.2b, we list flux density derived from a circular Gaussian fit with fixed FWHM of 0.88$''$. The peak flux density measurement is 12.5$\pm$3.5 $\mu$Jy. }}
\tablefoottext{c}{GN20 also has secure detections in SPIRE 350 and 500 $\mu$m bands \citep{magdis11} and in the SCUBA 850 $\mu$m band \citep{pope06,daddi09b}.}
}
\end{table*}

\subsection{The infrared properties}
\subsubsection{Millimeter continuum emission}

Figure~\ref{fig1} shows evidence for  3.3 mm continuum emission. Averaging the spectrum in the uv space for velocities outside of CO(4-3) line emission range, we find a bright source at the position of GN20, and 3.3~mm continuum detections with significance of 2.5$\sigma$ and 3.3$\sigma$ for GN20.2a and GN20.2b (Fig.~\ref{fig2}), respectively. The flux densities are measured by fitting Gaussian functions to the maps using sizes estimated from the CO imaging and are summarized in Table~\ref{tbl2}. The 3.3~mm flux density of GN20 is consistent with the lower S/N measurement reported in \citet{daddi09b}, while for GN20.2a and GN20.2b, they are presented here for the first time.

In addition, we detect 1.2 mm emission of GN20, GN20.2a, and GN20.2b in PdBI maps with significant S/N level (see Riechers et al., in prep.). Measured fluxes are in the range of 3.3--8.5 mJy (Table~2) and are used in the remainder of the paper. We also make use of 2.2 mm continuum data of these three galaxies, which have been reported in \citet{carilli10}. Similar to the measure of CO(4-3) flux density, we used a circular Gaussian model with a fixed FWHM, same as the one used for CO(4-3) map, to fit the 1.2 and 2.2 mm images for each galaxy. For a consistency check, assuming an GN20-like spectral energy distribution \citep[SED;][]{magdis11}, we extrapolate the flux density of 1.2 mm to 2.2 mm and 3.3 mm for GN20.2a and GN20.2b, and find these predictions match with the 2.2 mm and 3.3 mm flux densities derived from PdBI maps. The derived mm flux densities for these galaxies are summarized in Table~\ref{tbl2}.

\begin{figure}[htbp]
\centering
\includegraphics[width=0.45\textwidth]{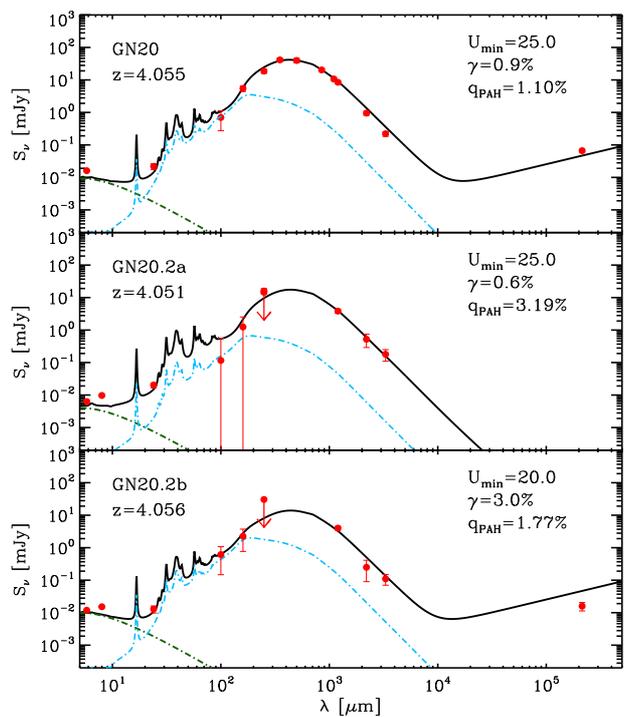}
\caption{Observed mid-IR to millimeter SED of GN20, GN20.2a, and GN20.2b, overlaid with the best-fit \citet{draine07} models. The black line is the DL07 model consisting of one stellar component and two dust components: diffuse ISM and photodissociation region.  Green dot-dashed and cyan dot-dashed lines show stellar component and ``PDR'' component, respectively. The best-fit parameters from DL07 models are listed within each panel. The arrows represent 5$\sigma$ upper limits.  \label{fig3}}
\end{figure}

\subsubsection{Dust masses, dust temperatures and SFRs}
Because they sample the rest-frame far-IR/submm bands at the Rayleigh-Jeans tail, the mm data have been demonstrated to have significant impact on the derivation of far-IR properties of high redshift galaxies \citep{magdis12a}, particularly crucial to reduce the uncertainties of dust mass estimates. Combining the {\it Spitzer} IRAC, MIPS photometry \citep{daddi09b} with the {\it Herschel} PACS 100 $\mu$m, 160 $\mu$m and SPIRE 250 $\mu$m photometry (for GN20.2a and GN20.2b, the 250 $\mu$m data were not used for SED fitting because the 250 $\mu$m photometry is blended between these two galaxies) and following \citet{magdis12a}, we derive the dust mass of the galaxies by fitting the IR SED using the \citet[][hereafter DL07]{draine07} dust models. The DL07 model describes the interstellar dust as a mixture of carbonaceous and amorphous silicate grains, with a size distribution mimicking the Milky Way extinction curve.  The fraction of dust mass in the form of polycyclic aromatic hydrocarbon (PAH) grains (with $<$ 10$^3$ carbon atoms) is parametrized by the PAH index, $q_{\rm PAH}$. The PAH abundance has an important effect in shaping the galaxy SED at short wavelengths. According to the DL07 model, the spectrum of a galaxy can be described by a linear combination of one stellar component  approximated by a blackbody with a color temperature of 5000~K and two dust components \citep[see][]{magdis12a}. The majority (i.e., $1-\gamma$ with $0.0<\gamma<0.3$) of the dust grains are located in the diffuse ISM and heated by a radiation field contributed by many stars with a minimum intensity $U_{\rm min}$ ($U$ is a dimensionless factor normalized to the local ISM), and the rest dust grains are exposed to a power-law radiation field ranging from $U_{\rm min}$ to $U_{\rm max}$, associated with photodissociation regions (PDRs). Following the prescriptions of DL07, we fit the mid-IR to millimeter data points for each galaxy. For GN20, we use the updated 3.3 mm flux density for SED fitting and find a consistent dust mass with the estimates in \citet{magdis11,magdis12a}. In order to more accurately derive far-IR properties, we also add 1.4 GHz measurements (Table~\ref{tbl2}; Owen et al. 2014, in preparation) to the SED fitting, with the exception of GN20.2a, due to the presence of an AGN \citep{daddi09b}. The 1.4 GHz flux density of GN20.2b measured from our new deep radio map is revised to a factor of 2 lower than the value reported previously \citep{daddi09b,morrison10}. The best-fit models along with the observed SEDs for GN20, GN20.2a, and GN20.2b are shown in Fig.~\ref{fig3}, and the derived parameters are summarized in Table~\ref{tbl2}. 

The radiation field intensities $\langle U \rangle$ derived from the best-fit models are 27.2$_{-2.2}^{+2.6}$, 26.4$_{-14.1}^{+8.1}$, and 25.8$_{-12.5}^{+6.3}$ for GN20, GN20.2a and GN20.2b, respectively. These can be represented as dust temperatures if assuming that the dust is heated only by radiation. Integrating over the best-fitting template from 8 to 1000 $\mu$m, we derive $L_{\rm IR} = (1.86\pm0.09)\times10^{13}\ L_\odot$, $(8.0\pm0.7)\times10^{12}\ L_\odot$, and $(6.9\pm1.0)\times10^{12}\ L_\odot$ for GN20, GN20.2a, and GN20.2b, respectively. We find that the uncertainties of IR luminosity estimates are significantly reduced by adding 1.4 GHz measurements. To convert the observed IR luminosities into SFRs, we adopt a \citet{kennicutt98} conversion of SFR[$M_\odot$ yr$^{-1}$] = $L_{\rm IR}$[L$_\odot$]/10$^{10}$, appropriate for a \citet{chabrier03} IMF.  The SFR derived in this way is used for analysis in the remainder of the paper.

\subsubsection{The Radio-IR correlation}

The IR-radio correlation is usually quantified by the so-called $q$-parameter \citep{helou85}:
\begin{equation}
q = {\rm log}(\frac{L_{\rm IR}}{3.75\times10^{12}\ {\rm W}})-{\rm log}(\frac{L_{\rm 1.4 GHz}}{{\rm W\ Hz^{-1}}}).
\end{equation}
where $L_{\rm 1.4 GHz}$ is the $k$-corrected radio luminosity assuming $S_\nu \propto \ \nu^\alpha$ with $\alpha=-0.8$. Using the 1.4 GHz flux density measurements (Table~\ref{tbl2}; Owen et al., in prep.) and the IR luminosities derived above, we find values of $q$=2.41$\pm$0.07, 1.72$\pm$0.05, and 2.60$\pm$0.19 for GN20, GN20.2a, and GN20.2b, respectively.  The $q$ values of GN20 and GN20.2b are compatible with the local relation ($q$=2.64$\pm$0.02,  scatter: 0.26; \citealt{bell03}) and galaxies at $z\sim$1--2 \citep{sargent10}, suggesting that dust heating in these two galaxies originates predominantly from star formation. The mid-IR observations of GN20 reveal a significant power-law component, likely related to the presence of an obscured AGN \citep{riechers14}. However, the infrared SED fits to GN20 after correcting for the AGN fraction show that the AGN contribution to the total IR luminosity is only $\lesssim$15\% \citep{riechers14}. In contrast, GN20.2a falls below the range defined for star-forming galaxies \citep{yun01,bell03}, the relatively low $q$ value is suggestive of AGN activity boosting radio flux. In addition, we found that the $q$ value of GN20.2a is similar to that of the Cloverleaf quasar at $z$=2.55, which hosts a known radio-loud AGN \citep{beelen06,kayser90}.

\subsection{Multiwavelength counterparts}

\begin{figure*}[htbp]
\centering
\begin{minipage}[t]{0.3\textwidth}
\includegraphics[width=\textwidth]{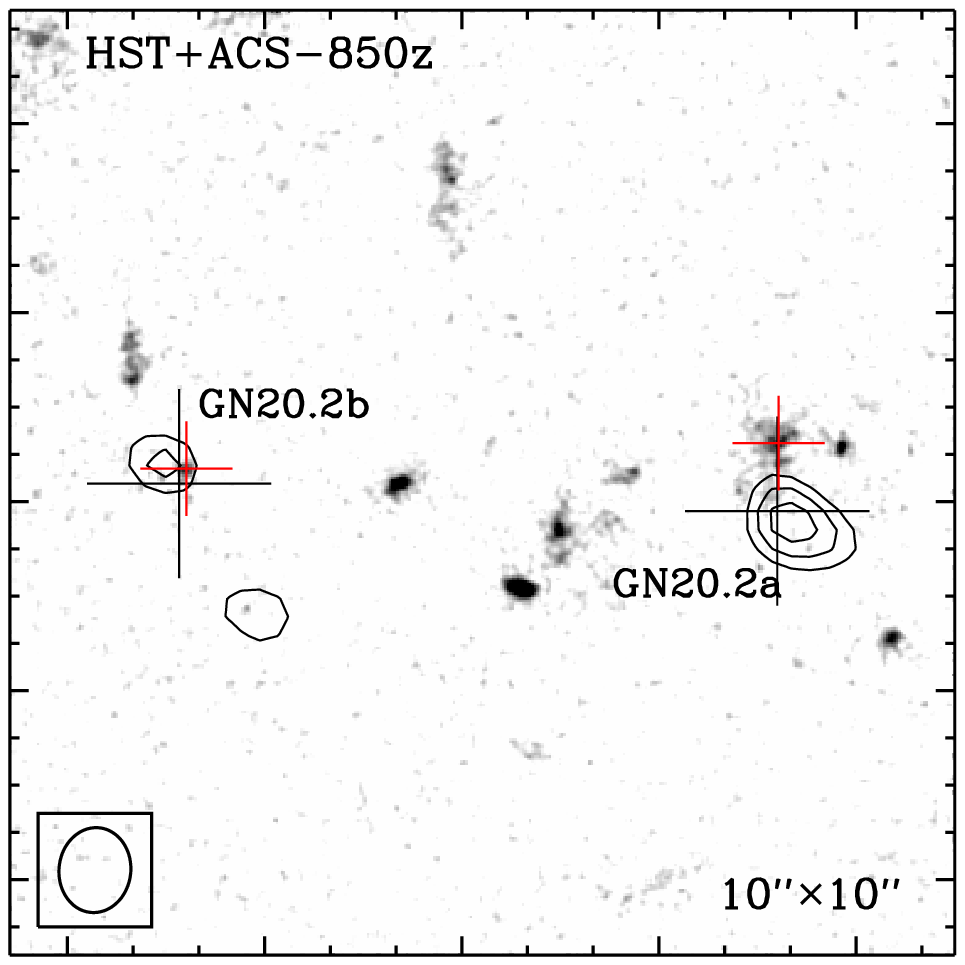}
\vskip-153pt
\hskip97pt
\includegraphics[width=0.35\textwidth]{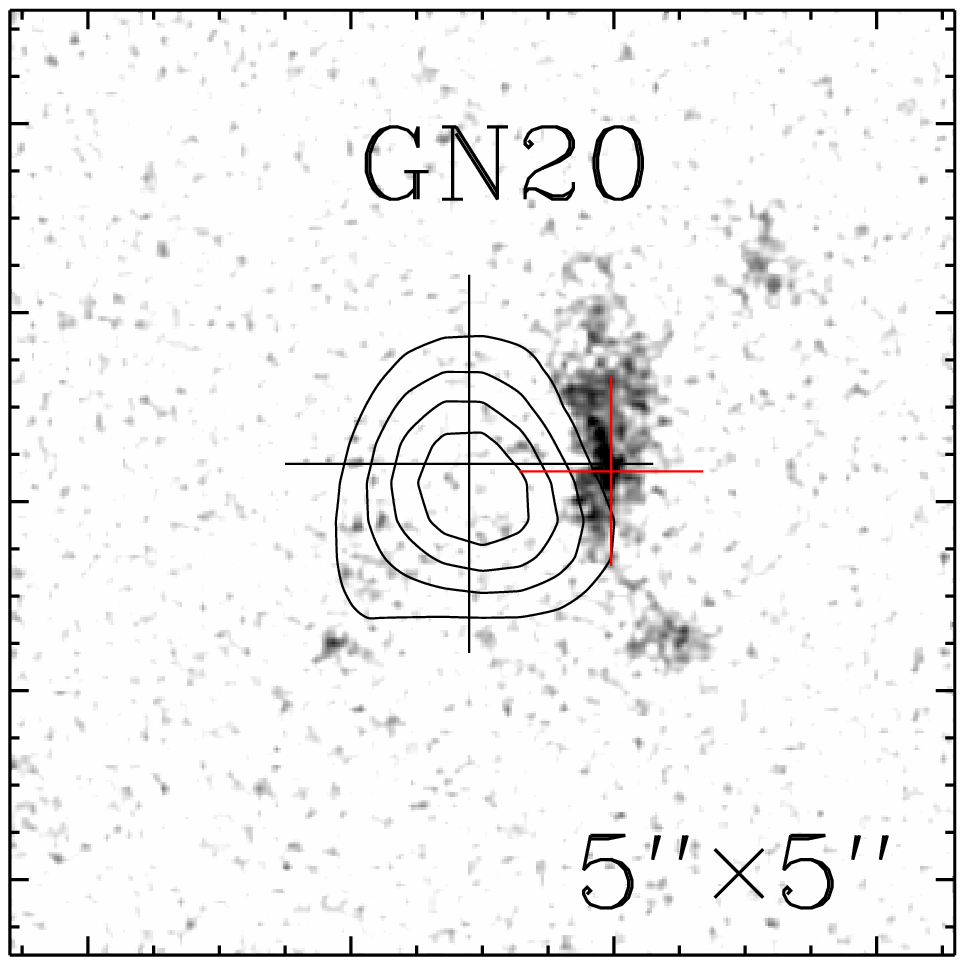}
\end{minipage}
\begin{minipage}[t]{0.3\textwidth}
\includegraphics[width=\textwidth]{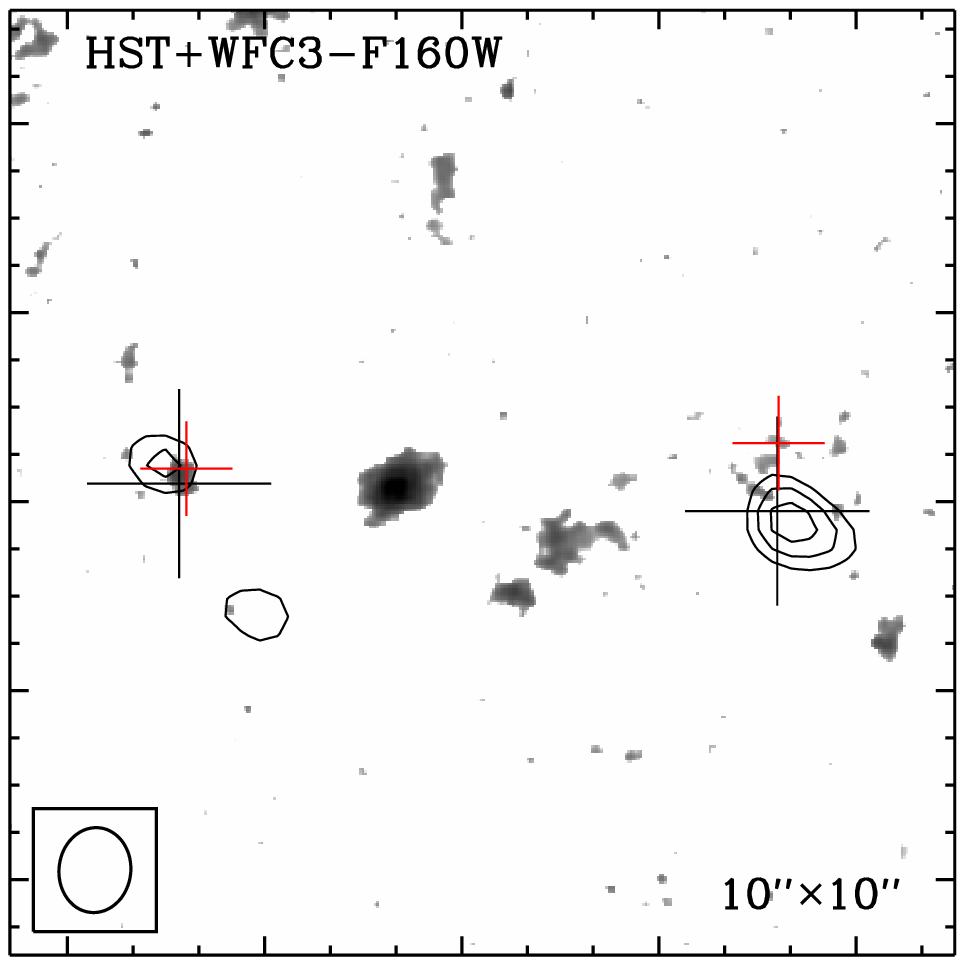}
\vskip-153pt
\hskip97pt
\includegraphics[width=0.35\textwidth]{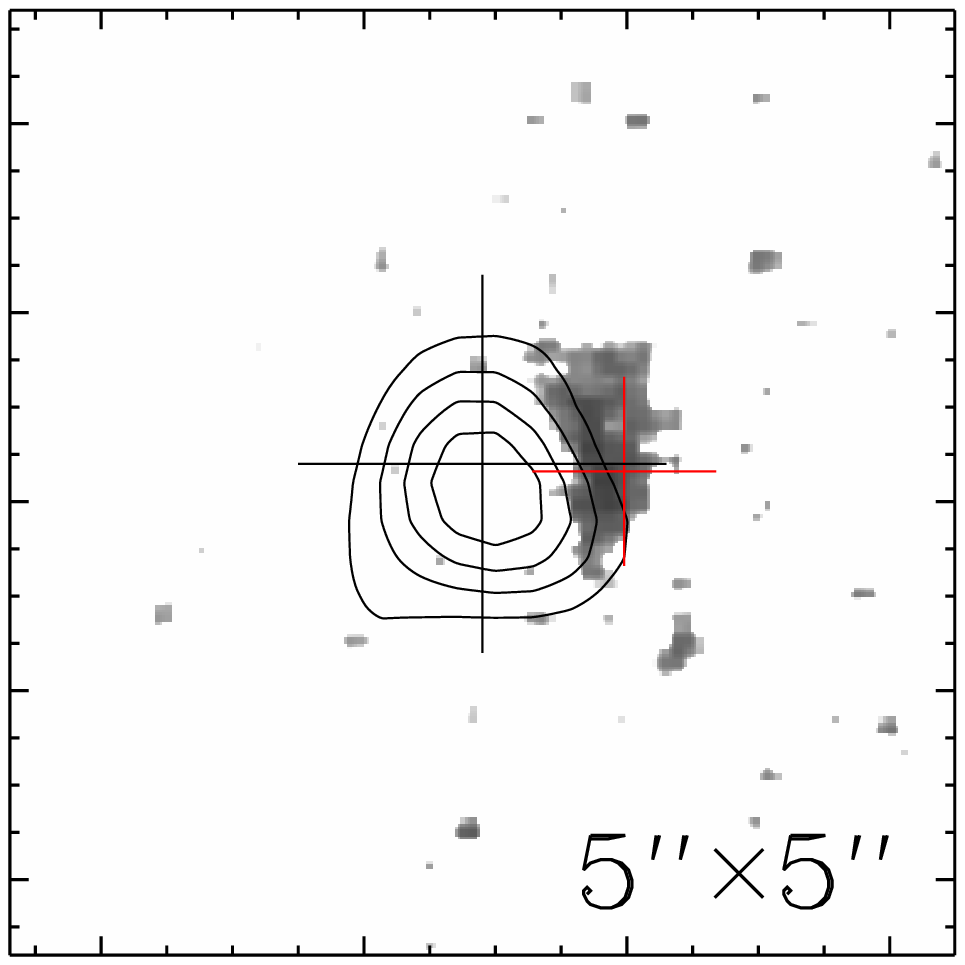}
\end{minipage}\\
\begin{minipage}[t]{0.3\textwidth}
\includegraphics[width=\textwidth]{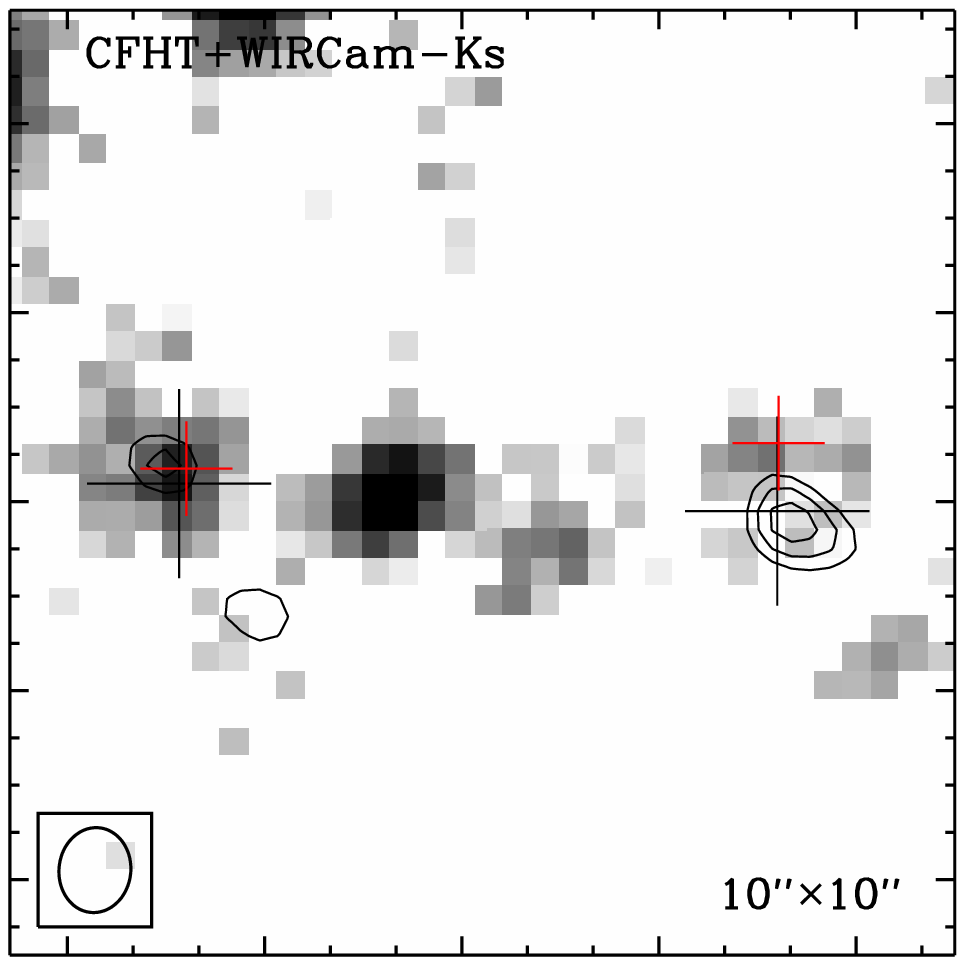}
\vskip-153pt
\hskip97pt
\includegraphics[width=0.35\textwidth]{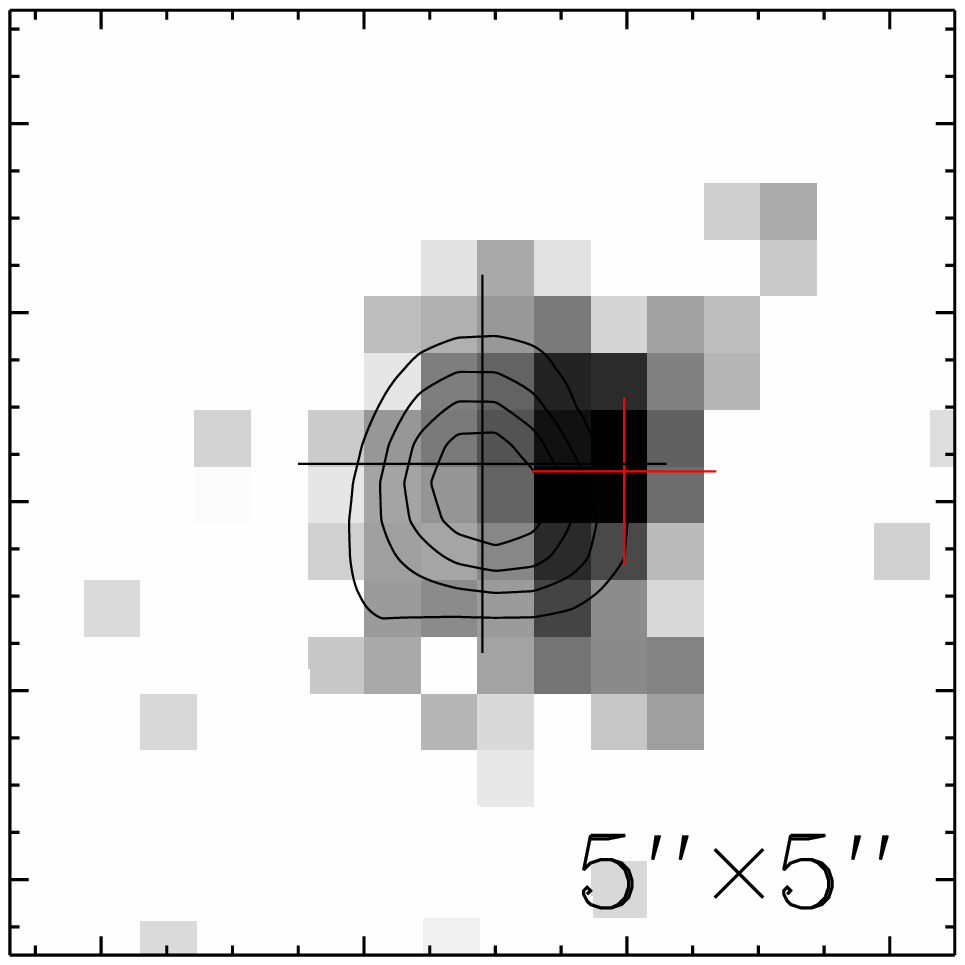}
\end{minipage}
\begin{minipage}[t]{0.3\textwidth}
\includegraphics[width=\textwidth]{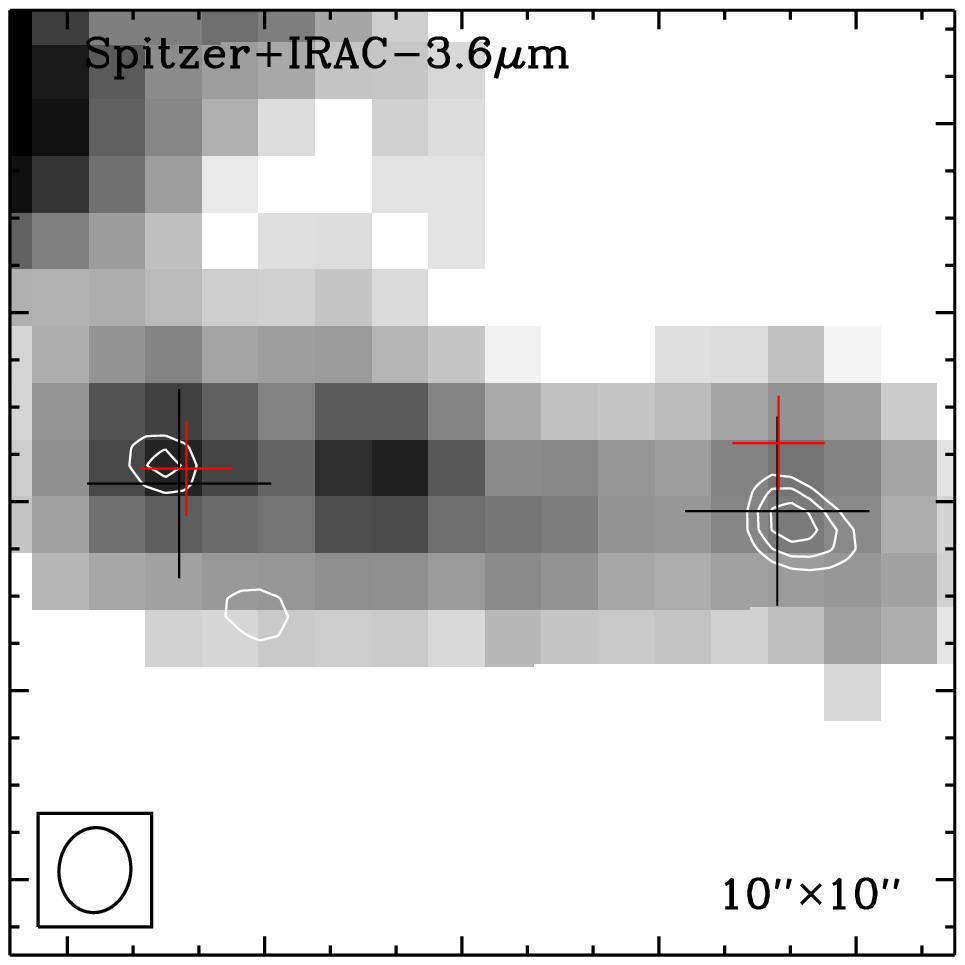}
\vskip-153pt
\hskip97pt
\includegraphics[width=0.35\textwidth]{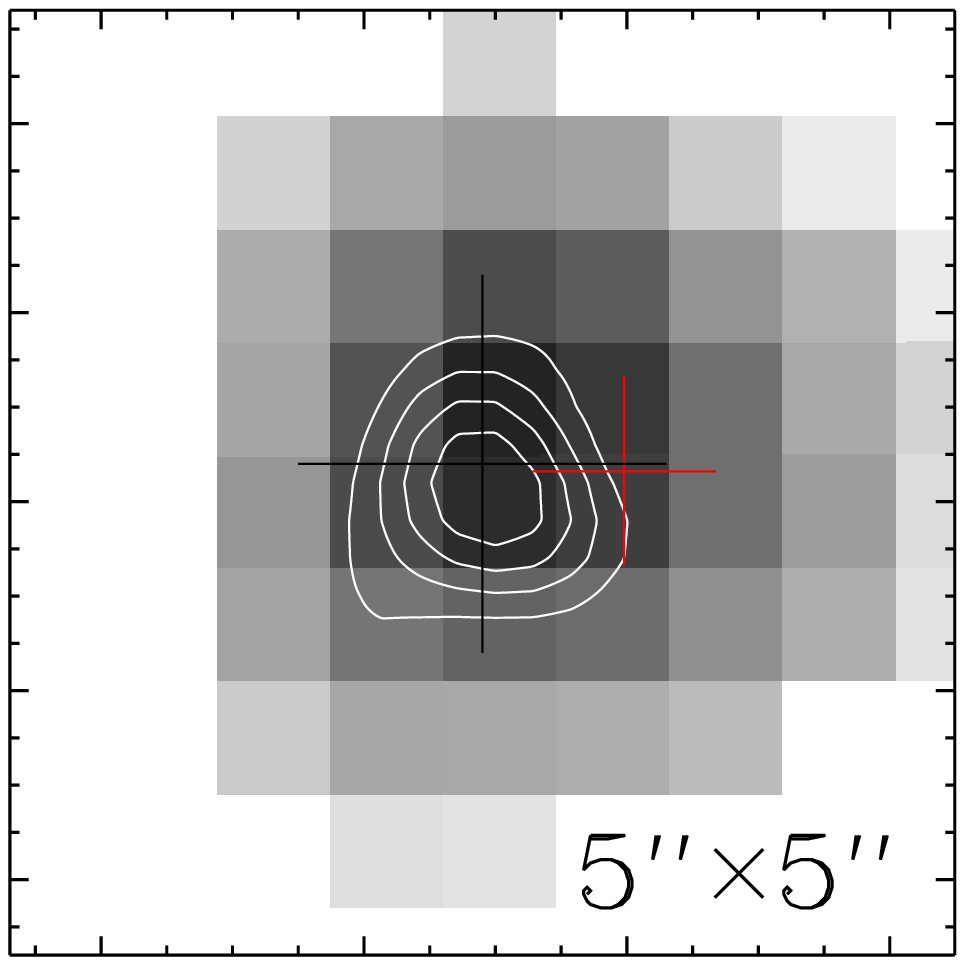}
\end{minipage}

\caption{PdBI CO(6-5)+continuum contours \citep{carilli10} at resolution of $0.90'' \times 0.76''$ for GN20 (inset), GN20.2a and GN20.2b overlaid on the ACS 850z-band (top left), WFC3 F160W-band (top right), WIRCam Ks-band (bottom left), and IRAC 3.6$\mu$m (bottom right) images. GN20 lies about 24$''$ to the northeast of GN20.2a. Black and red crosses show the positions of the 1.4 GHz counterparts (Owen et al., in prep.) and optical counterparts of GN20, GN20.2a and GN20.2b, respectively. Contour levels are shown in steps of 1$\sigma$ for GN20.2a and GN20.2b and 3$\sigma$ for GN20 starting at $\pm3\sigma$. The beam sizes of CO(6-5) are displayed in the left corner of each map. \label{counterparts}}
\end{figure*}

Figure~\ref{counterparts} shows the CO(6-5) contours of GN20, GN20.2a, and GN20.2b with resolution of  0.9$''$/6.1 kpc  overlaid on the HST+ACS 850z-band, the HST+WFC3 F160W-band, the CHFT+WIRCam K-band, and the Spitzer+IRAC 3.6 $\mu$m images. The WFC3 images are publicly available from the CANDELS survey \citep{grogin11,koekemoer11}. For GN20.2b, the multi-wavelength counterparts are found to be coincident with the CO source. We note that a nearby companion galaxy lies $\sim 2''$ to the west of GN20.2b, which is also likely to be a B-band dropout based on the ACS images shown in \citet{daddi09b}. For GN20.2a, we find a significant offset between CO position and optical counterpart, which has already been presented in \citet{hodge13} with higher resolution of CO(2-1) map. This is similar to the large offset found between CO and optical counterpart for GN20 \citep{carilli10}, revealing a substantial dust obscuration over a scale $\sim$ 10 kpc. The spectroscopic redshift ($z=4.059\pm0.007$; Daddi et al. 2009) derived for the optical counterpart is found to be very close to the CO redshift of GN20.2a, makes it likely that the HST galaxy to the northeast of the radio position is related to the CO emission (see Fig.~\ref{counterparts}). 

\subsection{Stellar mass and UV-based SFR estimates}\label{UV-SFR}
\begin{figure*}[htbp]
\centering
\scalebox{0.29}{\includegraphics*[30,145][565,535]{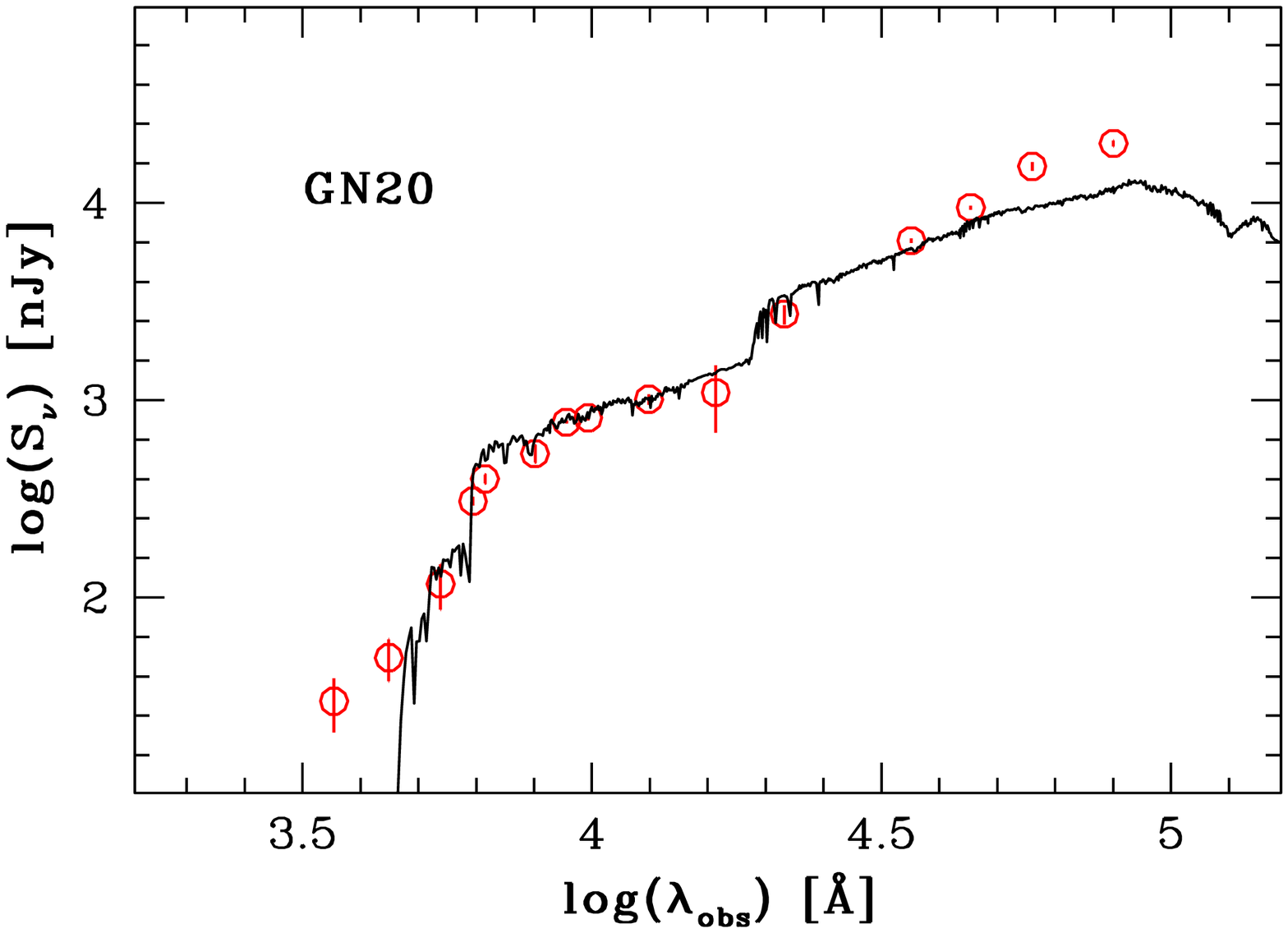}}
\scalebox{0.29}{\includegraphics*[30,145][565,535]{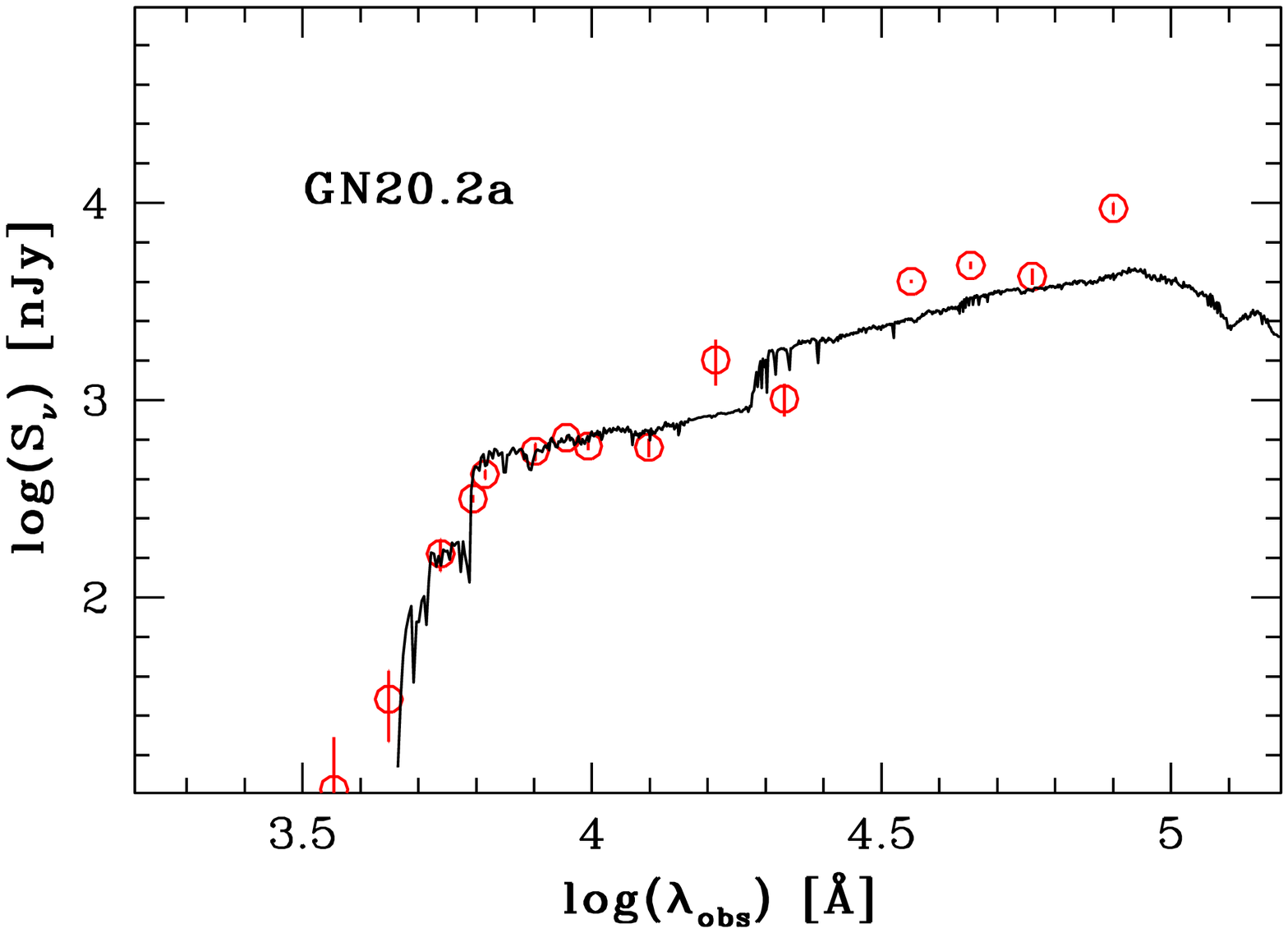}}
\scalebox{0.29}{\includegraphics*[30,145][565,535]{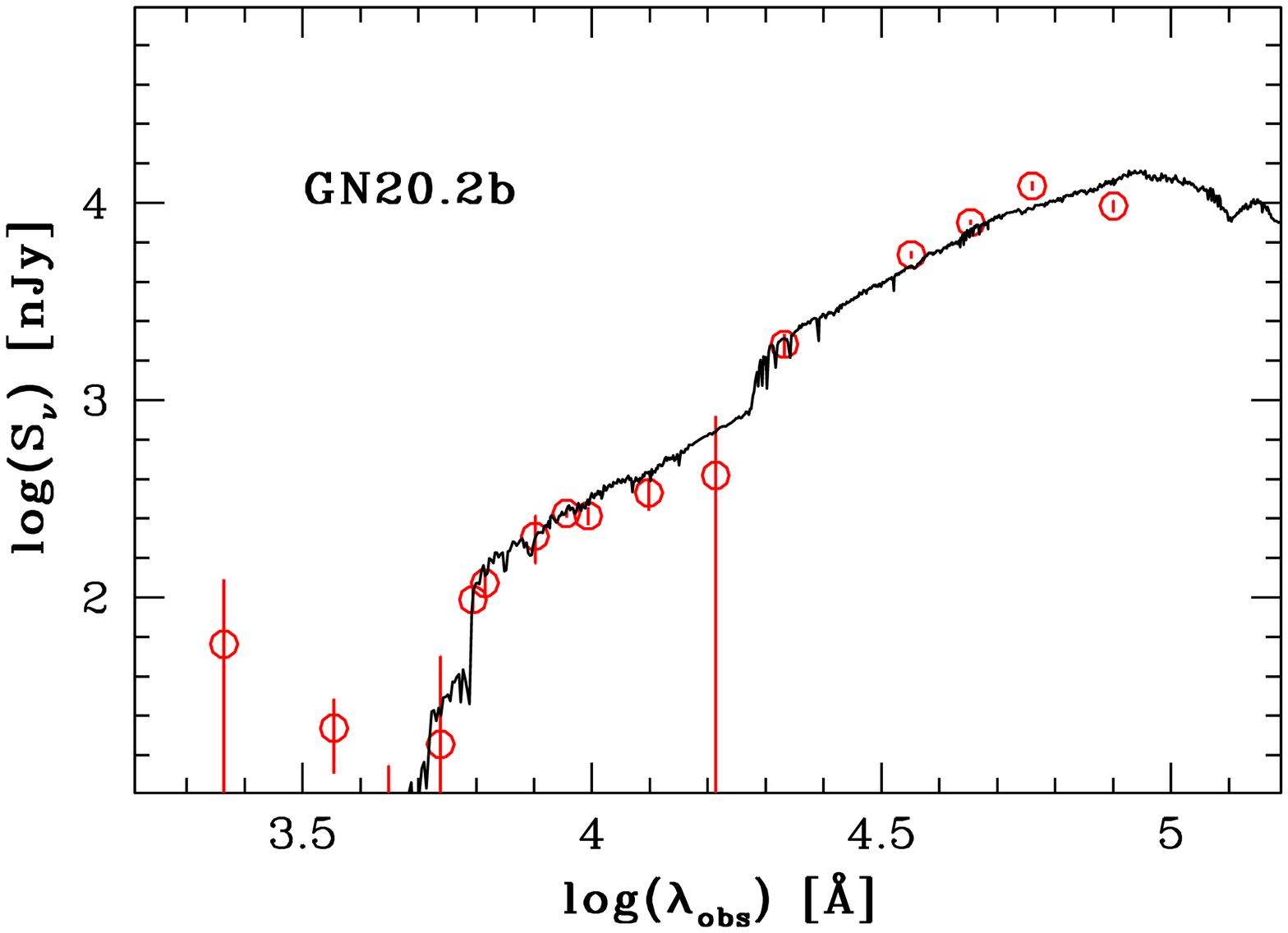}}
\caption{Observed optical to mid-IR SED of GN20 (left), GN20.2a (middle), and GN20.2b (right). Red circles with error bars represent the multi-wavelength photometry. The solid lines show the best-fitting templates from the libraries of \citet{bruzual03} at the derived CO redshift for each galaxy, assuming a constant SFH models with solar metallicity and using an extinction correction from \citet{calzetti00}. We adopt the templates with a constant SFH to derive stellar mass by fitting the photometry of U- to IRAC 4.5$\mu$m band.  \label{fig:sedfits}}
\end{figure*}

The photometric information we use to measure the physical properties of our galaxy sample is drawn from the GOODS-N multi-wavelength catalog presented in \citet{pannella14}. For the sake of clarity we give here only a brief summary of the catalog properties while referring the reader to \citet{pannella14} for a more detailed description. The extended GOODS-N catalog is a Ks-band selected multi-wavelength catalog spanning 20 passbands from GALEX NUV to IRAC 8 $\mu$m. PSF-matching corrections have been applied to account for the different angular resolution of the images. Aperture magnitudes (2$''$ diameter) are used to sample the galaxy SED. Finally, all derived properties were extrapolated to ``total'' masses using the ratio between the total (FLUX\_AUTO) and aperture flux in the K-band detection image. The catalog contains 53675 objects over the WIRCAM Ks image field of 900 arcmin$^2$ and down to an AB magnitude of 24.5, the 5$\sigma$ limiting magnitude of the image. The $UBVRIzJHK$ photometry, plus the photometry in IRAC bands are shown for GN20, GN20.2a, and GN20.2b in Fig.~\ref{fig:sedfits}. 

The stellar masses were estimated by fitting the multi-wavelength photometry of U- to IRAC 4.5 $\mu$m band to \citet{bruzual03} templates using FAST (Fitting and Assessment of Synthetic Templates; \citealt{kriek09}) through a $\chi^2$ minimization. We fix the spectroscopic redshift as given by the CO line identification. The IRAC 3.6 $\mu$m measurements were not used in the SED fitting, to avoid the possible flux contamination from H$\alpha$ emission, which happened to fall into this band. Given that all our galaxies are actively star forming, we adopt the templates with a constant star formation history (SFH), a large range of ages, a \citet{chabrier03} IMF, and allowing the metallicity to vary within a range of $Z$=0.2-2.5 solar. We used an extinction correction from \citet{calzetti00} with an optical extinction $A_{\rm V}$= 0$-$6 mag.  The typical uncertainty of stellar mass estimate is 0.2 dex.  For comparison, we also assume an exponentially declining SFH for the modeling and find the stellar masses are on average increased by 0.05 dex for each galaxy, while the UV-based SFRs are lowered by $\sim$ 0.5~dex. Therefore, the uncertainties of stellar mass estimates resulted from different assumptions of single-component SFH are relatively small for our sources. We also find that the stellar masses derived from the best-fitting model allowing a varied metallicity are not significantly different from the one assuming a solar metallicity. We adopt the modeling results derived by assuming a constant SFH for analysis in the remainder of the paper.

For GN20.2b, the stellar masses derived from the SED fitting up to K-band and to IRAC 4.5 $\mu$m band are differing by 0.2 dex. Given that the IRAC photometry for GN20.2b is likely to be blended with the companion source (see Fig.~\ref{counterparts}), we adopt the average of these two estimates, $M_{\star} = 1.1 \times 10^{11}\ M_\odot$, and an uncertainty of about 0.2 dex. Looking at the IRAC image of GN20.2a, we find that the IRAC peak is coincident with the northeast optical counterpart and close to the CO peak, and thus we take into account the IRAC 4.5 $\mu$m photometry in the SED fitting for the stellar mass derivation, yielding $M_{\star} = 3.8 \times 10^{10}\ M_\odot$. We stress that the stellar mass estimate of GN20.2a should be treated with caution, due to the possible contamination in the IRAC band emission from its nearby companion galaxy (see Fig.~\ref{counterparts}), and the possibility that the optical ``counterpart'' is a distinct unrelated galaxy from the CO-emitting dusty starburst galaxy. It is not easy to disentangle this issue given the current available data.

We estimated the UV-based SFR with the same SED fitting approach used to derive stellar masses. However, also to avoid possible uncertainties linked to blending in the IRAC bands, we only consider the photometry of U- to K-bands for the measure of UV-based SFR, as the observed K band corresponds to the rest-frame 4000 \AA \ Balmer break. Comparing the SFR estimated from dust-corrected UV luminosity with the one converted from IR luminosity, we find that the IR-based SFR estimates are generally larger than the UV-based ones for these three SMGs. The ratios of SFR$_{\rm IR}$/SFR$_{\rm UV}$ we derived are 13, 26, and 6 for GN20, GN20.2a, and GN20.2b, respectively.  As discussed above, the UV-based SFRs would be lowered by a factor of $\sim$3 if assuming an exponentially declining SFH, thus result in even higher SFR$_{\rm IR}$/SFR$_{\rm UV}$ ratios. The extremely large value of SFR$_{\rm IR}$/SFR$_{\rm UV}$ measured in GN20.2a is indicative of heavily obscured UV/optical emission at the positions of CO/radio emission, in agreement with the large offset observed between optical and CO peaks shown in Fig.~\ref{counterparts}. As discussed in \citet{daddi09b}, both the complex morphology and large ratio of IR/UV-derived SFRs may suggest that GN20.2a is undergoing a major merger with a B-band dropout companion ($\sim 0.7''$ west to GN20.2a; see Fig.~\ref{counterparts}).

\begin{figure}
\centering
\includegraphics[width=0.5\textwidth]{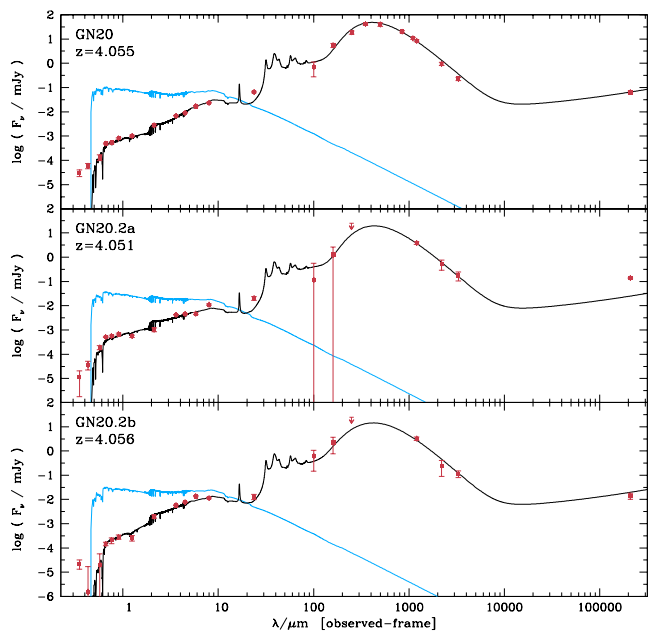}
\caption{Global UV to FIR fits to our galaxies based on MAGPHYS. The black line is the best-fit SED and the blue line is the corresponding unattenuated stellar emission.\label{fig:bets}}
\end{figure}

\subsection{Physical parameters from global UV to FIR SED fitting}

In order to verify the stability of our results and test against possible systematic effects affecting our modelling, we obtained simultaneous estimates of the stellar masses, SFRs, IR luminosities and dust masses using MAGPHYS \citep{dacunha08}. We use an updated version of MAGPHYS where the parameter priors are appropriate for high-redshift galaxies (da Cunha et al. 2014, in prep). This code includes a wide range of possible SFHs and dust properties, and fits the full SED consistently from the UV to the far-IR by requiring an energy balance between the UV/optical and far-IR output (Fig.~\ref{fig:bets}). 

We find in general excellent agreement with the results obtained from separate UV/optical and IR fitting described in previous sections. In all cases the 
SFRs are very close to what inferred from the IR luminosities applying a standard \citet{kennicutt98} conversion.
After correcting for the different $\kappa$ values ($\kappa$ is the dust absorption coefficient) adopted in MAGPHYS and  in the DL07 model, the differences of dust mass estimates for these three sources are within 25--30\%. For the stellar mass, except GN20.2b, the differences are also only ~20\%.

For GN20.2b, MAGPHYS provide two degenerate solutions for the stellar mass. One is consistent with our previous derivation (about $10^{11}M_\odot$) and a second one with much smaller values up to factors of 3--5.

\subsection{The CO-to-H$_2$ conversion factor}\label{alphaCO} 
Recently, a few attempts at the direct measurements of the CO luminosity to H$_2$ gas mass conversion factor for high redshift galaxies have been mainly based on dynamical modeling, gas-to-dust measurements, and radiative transfer modeling \citep[e.g.,][and references therein]{bolatto13,carilli13}. A pilot study performed by \citet{magdis11} showed that the gas-to-dust ratio method applied to GN20 leads to a conversion factor of $\alpha_{\rm CO}<1.0$, consistent with the dynamical estimate measured by \citet{carilli10}. Having derived the dust, stellar, and dynamical masses, we use two independent methods of dynamical modeling and gas-to-dust mass ratio to estimate $\alpha_{\rm CO}$. To avoid the uncertainties resulting from the extrapolation from high-order CO transitions, we adopt the low-$J$ ($J$=1, 2) CO luminosities measured by \citet{carilli10,carilli11}. The CO luminosities are summarized in Table~\ref{tbl3}.

\subsubsection{The dynamical mass method} 
	 
 Assuming that our SMGs are a mix of disc-like and virialized systems, we estimate the dynamical mass by taking the average of two different estimators: the isotropic virial estimator \citep[e.g.,][]{spitzer87,pettini01}
\begin{equation}
M_{\rm dyn}(r<r_{\rm 1/2}) = \frac{5 \sigma^2 r_{\rm 1/2}}{G}\label{eq2}
\end{equation}
where $r_{\rm 1/2}$ is the half-light radius, $\sigma$= $\Delta \upsilon_{\rm FWHM}$/2.35 is the one-dimensional velocity dispersion ($\Delta \upsilon_{\rm FWHM}$ is the line width), and $G$ is the gravitational constant, and the rotating disk estimator \citep{neri03}, corrected for $<{\rm sin^2}i> =2/3$ in mass:
\begin{equation}
M_{\rm dyn}(r<r_{\rm 1/2}) = 6 \times 10^4 \Delta \upsilon_{\rm FWHM}^2 r_{\rm 1/2}\label{eq3}
\end{equation}
This method has been widely applied to the dynamical mass estimate for SMGs \citep[e.g.,][]{tacconi08,bothwell13,hodge13}. Unlike the dynamical mass traced by HI emission line in local universe, the CO-based dynamical mass represents more compact star formation region \citep[e.g.,][]{solomon05,deblok14}. Adopting the FWHM of CO(4-3) line derived from our best-fitting double Gaussian models and the CO size measured from resolved CO(2-1) maps \citep{carilli10,hodge13}, the implied dynamical mass of GN20 within $r_{\rm 1/2}\sim$ 4 kpc is $(1.8\pm 0.2)\times 10^{11}$ M$_\odot$, of GN20.2a within $r_{\rm 1/2}\sim$ 2.5 kpc is $(1.9\pm0.9)\times 10^{11}$ M$_\odot$, and of GN20.2b within $r_{\rm 1/2}\sim$ 4 kpc is $(2.6\pm1.0)\times 10^{10}$ M$_\odot$. The uncertainties were estimated based on the uncertainties on the FWHM measures.  The dynamical mass estimates for GN20 and GN20.2a are consistent within uncertainties with the estimates by \citet{carilli10} and \citet{hodge12,hodge13}. For GN20.2b, our dynamical mass estimate is better constrained than the one measured by \citet{hodge13}, as the uncertainty in its fitted line width is significantly reduced.

\begin{table*}
\caption{Physical properties of GN20, GN20.2a and GN20.2b}
\label{tbl3}
\centering
\small\addtolength{\tabcolsep}{-4pt}
{\renewcommand{\arraystretch}{1.5}

\begin{tabular}{lccccccccccccc}

\hline\hline
Source & 
$M_\star$ & 
sSFR & 
$L'_{\rm CO[1-0]}$ & 
SFE & 
Z$_{\rm PP04}$ & 
$\delta_{\rm GDR}$& 
$\alpha_{\rm CO}$(GDR) & 
$M_{\rm dyn}$ &
$\alpha_{\rm CO}$(dyn.) &
 $\langle \alpha_{\rm CO} \rangle$ &
$M_{\rm gas}$ &
$f_{\rm gas}$  &
$q$ \\
 & 
($10^{10} M_{\odot}$) & 
(Gyr$^{-1}$) & 
(10$^{10}$ K km s$^{-1}$ pc$^2$)  & 
($l_0$) &
  &  
  & 
  &
 ($10^{11} M_\odot$) &
  & 
  & 
 ($10^{11} M_{\odot}$) &
  &
 \\
 (1) & 
 (2) & 
 (3) & 
 (4) & 
 (5) & 
 (6) & 
 (7) & 
 (8) & 
 (9) & 
 (10) &
 (11) &
 (12) &
 (13) &
 (14) \\
\hline

GN20  & 11 & 16.9 & $16\pm0.4$ & 116 & 8.8(9.2) & 77(27)  & 1.6$\pm$0.7 & $1.8\pm0.2$ & 1.0$\pm$0.2 &  1.3$\pm$0.4 & $2.1\pm0.6$ & 0.66$\pm$0.12 & 2.41$\pm$0.05 \\

GN20.2a  & 3.8 & 20.9 & $9.0\pm1.1$ & 90 & 8.6(9.0)  & 106(42) & 2.0$\pm$1.1 & $1.9\pm0.9$ & 2.8$\pm$1.6 &  2.4$\pm$1.3 & $2.2\pm1.2$ & 0.85$\pm$0.09  & 1.72$\pm$0.04 \\

GN20.2b  & 11  & 6.3 & $2.9\pm0.6$ & 238 & 8.9(9.2) & 53(27) & 2.8$\pm$1.5 &  ... & ... & 2.8$\pm$1.5 & $0.8\pm0.5$ & 0.42$\pm$0.18  & 2.60$\pm$0.13 \\ 

\hline
\end{tabular}
}
\tablefoot{{Column 1}: Name of the object.  {Column 2}: Stellar mass determined from synthetic template fitting of the U- to IRAC 4.5$\mu$m-band photometry (without 3.6 $\mu$m-band), assuming a constant SFH and a Chabrier (2003) IMF. The typical uncertainties are 0.2 dex. For GN20.2b, the stellar mass is estimated by taking an average of the results from SED fitting with and without 4.5 $\mu$m-band. {Column 3}: Specific star formation rate. sSFR = SFR/M$_\star$, with SFR derived from the bolometric IR luminosity. { Column 4}: For GN20, the CO(1-0) luminosity is measured by \citet{carilli10}. For GN20.2a and GN20.2b, we convert the CO(2-1) luminosity measured by \citet{carilli11} to CO(1-0) luminosities by assuming $r_{21}$=0.84 \citep{bothwell13}. { Column 5}: SFE = $L_{\rm IR}/L'_{\rm CO}$ in units of $l_0$=$L_\odot$ (K km s$^{-1}$ pc$^2$)$^{-1}$. { Column 6}: Metallicity estimated based on the FMR (values outside the parenthesis) and mass-metallicity (values in the parenthesis) relation applied to the present-day elliptical galaxies. We have corrected for the metallicity scale from KD02 to PP04 for the FMR by adopting the prescriptions in \citet{kewley08}. { Column 7}: Gas-to-dust ratio derived from the relation of \citet{magdis12a} which relates the $\delta_{\rm GDR}$ to metallicity. The values in the parenthesis are inferred assuming a metallicity estimated from the mass-metallicity relation of present-day elliptical galaxies, while the values outside the parenthesis are from the FMR relation. { Column 8}: CO-to-H$_2$ conversion factor in  units of $M_{\odot}$ (K km s$^{-1}$ pc$^2$)$^{-1}$, inferred from $\alpha_{\rm CO}=M_{\rm gas}/L'_{\rm CO[1-0]}$, with $M_{\rm gas}$ = $M_{\rm dust}\times \delta_{\rm GDR}$.  { Column 9}:  dynamical masses within the half-light radius. For GN20.2b, we do not derive an estimate of dynamical mass, as the dynamical mass estimated in this way is found to be smaller than stellar mass. { Column 10}: CO-to-H$_2$ conversion factor estimated based on the dynamical method.  { Column 11}: The average of CO-to-H$_2$ conversion factor derived from gas-to-dust ratio method and dynamical modeling. { Column 12}: Molecular gas mass measured from $L'_{\rm CO[1-0]}$ and  $\langle \alpha_{\rm CO} \rangle$. { Column 13}: Molecular gas fraction: $f_{\rm gas} = M_{\rm gas}/(M_{\rm gas}+M_\star)$.The uncertainties are mainly dominated by the uncertainties in molecular gas estimates. { Column 14}: IR-radio correlation parameter.}

\end{table*}

To estimate the gas mass from the dynamical modeling, we follow the method in \citet{daddi10}
\begin{equation}
M(r<r_{\rm 1/2})=0.5\times(M_\star+M_{\rm gas})+M_{\rm dark}(r<r_{\rm 1/2})\label{eq4}
\end{equation}
where the dynamical mass within the half-light radius is composed of the half of the total amount of stellar mass and gas mass, and the amount of dark matter within $r_{\rm 1/2}$. Subtracting the stellar mass, including a 25\% dark matter contribution, which is a typical value adopted for $z\sim 1-2$ normal galaxies \citep{daddi10} and $z\sim 2$ SMGs \citep{tacconi08}, we find $\alpha_{\rm CO} = 1.0\pm0.2$ and $2.8\pm1.6$ for GN20 and GN20.2a, respectively. The derived $\alpha_{\rm CO}$ for GN20 is consistent with previous estimates \citep{carilli10,hodge12}.  For GN20.2b, the dynamical mass is found to be much smaller than our preferred estimate of the stellar mass, suggesting that our method may possibly significantly underestimate the dynamical mass. If we adopt the rotating disk estimator and make an assumption for the extreme case that GN20.2b is composed of stellar mass and dark matter only, the maximum inclination angle derived for this galaxy would be about 20$\degr$, close to face-on. This is consistent with the narrow CO line width that we observed.  On the other hand,  the alternative lower stellar mass value suggested by MAGPHYS for this object is compatible with the inferred dynamical mass. Nevertheless, given all the uncertainties,  no inference on  dynamical mass and $\alpha_{\rm CO}$ for GN20.2b can be directly obtained in this way. 

\subsubsection{The gas-to-dust ratio method}\label{gdr}
Some recent studies have revealed a tight correlation between the gas-to-dust ratio ($\delta_{\rm GDR}$) and the gas-phase oxygen abundance, with $\delta_{\rm GDR}$ decreasing for more metal-rich galaxies \citep[e.g.,][]{leroy11,magdis11}. Under the assumption that the local observed $\delta_{\rm GDR}-Z$ relation is valid at high redshift, \citet{magdis12a}  derived indirect estimates of $\alpha_{\rm CO}$ for a sample of star-forming galaxies at $z\sim2$ and found that the variation tendency of $\alpha_{\rm CO}$ is in line with previous studies \citep{leroy11}. Following the prescription in \citet{magdis12a}, we attempt to apply this approach to the three galaxies in our sample to estimate $\alpha_{\rm CO}$. 

Having obtained the dust masses for our sources, we need the measurements of their metallicities, for which we have to rely on indirect indicators. The fundamental metallicity relation (FMR) of \citet{mannucci10} that relates the metallicity to both SFR and $M_\star$ is one of the methods that is commonly used for measuring metallicity. However, the FMR relation is valid only up to $z\sim 2.5$, and thus we also apply the mass-metallicity relation of present-day elliptical galaxies \citep[e.g.,][]{calura09}, assuming that the high redshift strongly star-forming galaxies are the progenitors of the local ellipticals, given that the large SFR ($\sim 690-1860\ M_{\odot}$ yr$^{-1}$) of our sources could be due to a final burst of star formation triggered by a major merger, and the galaxy will eventually evolve into a massive elliptical with unchanged mass and metallicity once there are no further star formation activities. The metallicity estimates based on this scenario and the FMR relation lead to 12+log[O/H] = 8.8-9.2 for GN20 (agree with the values in \citealt{magdis11}), 8.6-9.0 for  GN20.2a, and 8.9-9.2 for GN20.2b. 

With the derived metallicity, we estimate the $\delta_{\rm GDR}$ following the relation of \citet{magdis12a} which relates the $\delta_{\rm GDR}$ to metallicity, i.e.,
\begin{equation}
{\rm log}\delta_{\rm GDR} =(10.54\pm1.0)-(0.99\pm0.12)\times({\rm 12+1og[O/H]})
\end{equation}
The fit to this relation indicates a value for $\delta_{\rm GDR}$ of $\sim$ 77(27) for GN20, $\sim$ 106(42) for GN20.2a, and $\sim$ 53(27) for GN20.2b, assuming a metallicity estimated based on the FMR relation (mass-metallicity relation of present-day elliptical galaxies). With the gas-to-dust ratios, we can determine the $M_{\rm gas}$ and estimate $\alpha_{\rm CO}$ from the equation $\alpha_{\rm CO}$ = $L'_{\rm CO}/M_{\rm gas}$. Here we have assumed that at high redshift $M_{\rm H_2}\gg M_{\rm HI}$ and therefore that $M_{\rm gas}\backsimeq M_{\rm H_2}$. This assumption is based on the observed high surface densities, above the characteristic threshold for HI saturation and thus most of the gas is in molecular form \citep[e.g.,][]{bigiel08,obreschkow09}. The $\alpha_{\rm CO}$ derived based on gas-to-dust method are $1.6\pm0.7$, $2.0\pm1.1$, and $2.8\pm1.5$ for GN20, GN20.2a, and GN20.2b, respectively.  We note that our $\alpha_{\rm CO}$ estimate for GN20 is larger than previously found by \citet{magdis11} with the same method, which is due to the revised estimates of dust and stellar mass used in our study. We caution that these $\alpha_{\rm CO}$ estimates are systematically uncertain due to the large uncertainties of metallicity estimates.  

\subsubsection{Comparison of the CO-to-H$_2$ conversion factor}

Comparing the $\alpha_{\rm CO}$ estimates summarized in Table~\ref{tbl3}, we find that the $\alpha_{\rm CO}$ determined based on the metallicity-dependent gas-to-dust ratio method are consistent within uncertainties with those measured from the dynamical modeling for each galaxy.  The average $\alpha_{\rm CO}$ are $1.3\pm0.4$, $2.4\pm1.3$, and $2.8\pm1.5$ for GN20, GN20.2a, and GN20.2b, respectively, which are found to be between the typical value ($\sim$0.8) determined for ULIRGs and the value ($\sim$4) appropriate for the Milky Way.  However, these values might be well below the value appropriate for normal galaxies at those epochs, as the $\alpha_{\rm CO}$ of $z>4$ normal galaxies could be much higher than the  Milky Way value  \citep[e.g.,][]{genzel12,bolatto13,carilli13,tan13}. We caution that the uncertainty of $\alpha_{\rm CO}$ estimates derived from each method is significant. For the dynamical analysis, the main uncertainties are the crude estimates of dynamical mass based on marginally resolved imaging data and the stellar mass based on SED fitting, which is caused by the extreme dust obscuration and the potentially complex SFHs,  while for the gas-to-dust method, the large uncertainties might result from the dust model assumed and the uncertainty on metallicity estimate. The consistency of $\alpha_{\rm CO}$ estimates derived from two independent approaches further confirm the reliability of the gas-to-dust ratio method.  

\subsection{The specific star formation rates and star formation efficiency}\label{ssfr}

With the stellar masses and star formation rates derived above, we estimate sSFR of 16.9 Gyr$^{-1}$, 20.9 Gyr$^{-1}$, and 6.3 Gyr$^{-1}$ for GN20, GN20.2a, and GN20.2b, respectively. Both GN20 and GN20.2a exhibit larger sSFRs with a factor of  $\sim$ 6  than the average of equally massive galaxies on the SFR-M$_\star$ main-sequence at $z\sim 4$ (Sargent et al. 2013), while GN20.2b shows a sSFR-excess of 2.4 (which could be larger though, in case of the lower-mass MAGPHYS solution is adopted instead). The large sSFR-excess for GN20 and GN20.2a, similar to the $z\sim 2-3$ SMGs, have been revealed in \citet{daddi09b}, suggesting that these galaxies are undergoing starburst events with short duty cycles, likely triggered by mergers. 

The ratio between $L_{\rm IR}$ and $L'_{\rm CO}$ (i.e., SFE, in units of $L_\odot$ (K km s$^{-1}$ pc$^2$)$^{-1}$) can be used as a measure of the efficiency with which molecular gas is converted into stars. Given that the total available reservoir of gas within a galaxy could be better traced by low-J CO emission, we use directly observed CO(1-0) luminosity or the one extrapolated from CO(2-1) observations for SFE derivation to avoid the uncertainties in the underlying gas excitation. We find ratios of $\sim$90--238 for these three galaxies, comparable to those derived for other high-redshift SMGs \citep{ivison11,bothwell13}, and higher than in normal star-forming galaxies ($\sim$84$\pm$12; Daddi et al. 2010), but slightly lower than in local ULIRGs ($\sim$250$\pm$30; Solomon et al. 1997). 

\subsection{Gas fractions, depletion timescales, and the evolutionary status}\label{fraction}

Table~\ref{tbl3} lists the molecular gas masses for our three galaxies, derived based on the average $\alpha_{\rm CO}$ inferred in Section~\ref{alphaCO} and the measured $L'_{\rm CO}$. Combining the gas mass with the stellar mass estimates, we find molecular gas fraction of $f_{\rm gas}=M_{\rm gas}/(M_{\rm gas}+M_\star)$ = 0.66$\pm$0.12, 0.85$\pm$0.09, and 0.42$\pm$0.18 for GN20, GN20.2a, and GN20.2b, respectively. These values are comparable to other SMGs at $z=2-4$ \citep{bothwell13} and high-$z$ massive, gas-rich star-forming galaxies \citep[e.g.,][]{daddi10,tacconi10}. \citet{bothwell13} compared the gas fraction of SMGs to that of local LIRGs and found an increase of gas fraction up to $z\sim 2$ followed by a flattening toward higher redshift. This is similar to the redshift evolution trend of gas fraction revealed in normal star-forming galaxies \citep{magdis12b,tan13}, despite that SMGs have typically an order of magnitude higher SFRs than normal galaxies. The 2 star formation mode (2-SFM) predictions by \citet{sargent13} have also revealed that the average gas fraction of starbursts is only slightly lower than that of main sequence galaxies. 

Dividing the stellar masses by the SFRs, we estimate the stellar mass building-up timescale of $\tau_{\rm build} \sim$ 59 Myr, 48 Myr, and 159 Myr for GN20, GN20.2a, and GN20.2b, respectively, if assuming that these galaxies have sustained their current SFRs continuously. Similarly, assuming that the SFRs continues at the current rate and neglecting the effect of feedback, the minimum time for exhausting the molecular gas reservoir can be given by the gas depletion timescale, $\tau_{\rm gas} = M$(H$_2$)/SFR, which we find to be $\sim$113 Myr, $\sim$ 275 Myr, and $\sim$116 Myr for GN20, GN20.2a, and GN20.2b, respectively. These timescales are relatively short than in massive, normal galaxies ($\sim$0.4--0.9 Gyr; e.g.,  Daddi et al. 2010; Tacconi et al. 2010), indicating a rapid star formation mode with intense burst for SMGs, probably undergoing major mergers or rapid cold accretion \citep[e.g.,][]{narayanan10,dave10}. In addition, the extremely high gas surface density revealed in GN20.2a may indicate a triggering mechanism of major merger, while the surface density of GN20.2b is found to be comparable to normal star-forming galaxies \citep{hodge13}. If we define starburst phase to be the fraction of stars have already been formed over the available gas reservoir, simply calculated as $M_\star/(M_{\rm gas}+M_\star)$, taking these numbers at face value we would find that the relatively young stellar age and long gas-consumption timescale of GN20.2a place it at an early stage, $\sim$15\%$\pm$9\% of the way through its starburst, while GN20 and GN20.2b have experienced $\sim$35\%$\pm$14\% and $\sim$60\%$\pm$35\% of their starburst phase, respectively. The CO excitation analysis has also revealed different merging states for these three galaxies, of which GN20.2b display lowest excitation \citep{carilli10,hodge13}.

\begin{table}
\caption{Derived properties of GN20, GN20.2a, and GN20.2b}
\label{tbl4}
\centering
\small
\addtolength{\tabcolsep}{-5.5pt}
{\renewcommand{\arraystretch}{1.5}

\begin{tabular}{lcccccc}
\hline\hline

Source &
$\frac{{\rm sSFR}}{{\rm sSFR}_{\rm MS}}$ &
$\frac{{\rm SFR}_{\rm IR}}{{\rm SFR}_{\rm UV}}$ &
$\tau_{\rm gas}$ &
$\tau_{\rm build}$ &
Source size$\tablefootmark{a}$ &
$\Sigma_{\rm gas}\tablefootmark{b}$ \\
 &
 &
 &
(Myr) &
(Myr) &
(kpc) &
(M$_\odot$ pc$^{-2}$) \\

\hline
GN20 & 6.4 & 13 & 113 & 59 & $\sim$~8 &  $\sim$2400 \\
GN20.2a & 6.3 & 26 & 275 & 48 & $\sim$~5$\times$3 & $\sim$3900$\times$(sin $i$)($\alpha_{\rm CO}$/0.8) \\
GN20.2b & 2.4 & 6 & 116 & 159 & $\sim$~8$\times$5 & $\sim$530$\times$(sin $i$)($\alpha_{\rm CO}$/0.8) \\
\hline
\end{tabular}
}
\tablefoot{
\tablefoottext{a}{Deconvolved CO(2-1) size (Gaussian FWHM). From \citet{carilli10} and \citet{hodge13}.}
\tablefoottext{b}{Average gas surface density. From \citet{carilli10} and \citet{hodge13}.}
}
\end{table}

\section{What is the nature of GN20, GN20.2a and GN20.2b?}\label{nature}

We have presented multiwavelength properties of GN20, GN20.2a and GN20.2b, including UV/optical, far-IR, and mm photometry and molecular gas content. Table~\ref{tbl4} lists the ratios of sSFR/sSFR$_{\rm MS}$ and SFR$_{\rm IR}$/SFR$_{\rm UV}$ for these three SMGs. The large sSFR-excess observed for GN20 and GN20.2a suggest that these two galaxies are starbursting outliers above the main sequence \citep{rodighiero11,sargent13}. Although GN20.2b  might situate within the MS scatter with sSFR/sSFR$_{\rm MS}\sim$2.4, the large value of SFR$_{\rm IR}$/SFR$_{\rm UV}$ indicates that a large fraction of UV emission from this galaxy is obscured, pointing also toward a population of starburst galaxy \citep{daddi07b,daddi10}.  For GN20 and GN20.2a, the large offsets between the CO positions and the optical counterparts show clear evidence of extreme obscuration, consistent with the heavy dust extinction observed in the UV emission. In addition, the high gas surface densities derived for GN20 and GN20.2a suggest that the star formation is dominated by the compact star-forming sites \citep[Table~\ref{tbl4};][]{carilli10,hodge12,hodge13}. As discussed in Sect.~\ref{fraction}, the gas-consumption time-scales (see Table~\ref{tbl4}) of these three SMGs are found to be significantly shorter than those of normal galaxies, indicative of a more rapid star formation mode. The extreme SFRs ($\gtrsim$700 M$_\odot$ yr$^{-1}$) and high radiation field intensity ($\langle U \rangle>$25) observed for our SMGs would also favor the scenario of intense starbursts, as suggested by Magdis et al. (2011) \citep[see also][]{magnelli12b}. In addition, our three SMGs appear to be dynamically distinct from the normal disk galaxies, since these galaxies are clearly separated from disk galaxies in the velocity-size plane, which can be used as a tool to constrain the angular momentum properties of galaxies \citep{courteau97,bouche07}. The low orbital angular momentum observed in our SMGs might be caused by a recent or ongoing merger. 

Two main modes are typically considered for triggering star formation in high redshift SMGs: major mergers \citep{narayanan10} and secular mode with smooth mass infall from the intergalactic medium and along the cosmic web \citep{dekel09,dave10}. However, it is still a matter of debate to identify which mode of star formation is responsible for the majority of the SMG population. High-resolution CO imaging of local ULIRGs and $z\sim 2$ SMGs have revealed large reservoirs of molecular gas concentrated in the galaxy nuclei with disturbed CO kinematics/morphologies, suggesting a merger-driven mechanism for the star formation activity \citep{downes98,tacconi08,engel10}. However, the resolved CO imaging of GN20 shows  ordered rotation and an extended gas distribution \citep{carilli10, hodge12}, which could be suggestive of a rotating disk. On the other hand, hydrodynamical simulations of gas-rich mergers have shown that the merger remnant gas can cool quickly and produce an extended star-forming disk \citep{springel05,robertson06,robertson08}. 

GN20.2a has a smaller CO size than GN20 and GN20.2b, but exhibits a much larger line width, implying a much deeper gravitational potential well, as CO line emission traces the kinematics of the potential well where the molecular gas lies.  Comparing the $L'_{\rm CO}$ between the one observed directly and the one predicted from the $L'_{\rm CO}$-FWHM relation discovered in \citet{bothwell13}, we find the observed $L'_{\rm CO}$ is in good agreement with the prediction for all our three SMGs, indicating that the dynamics within the gas emission region is dominated by the baryons. Given that the stellar mass of GN20.2a is the smallest one among these three galaxies, we infer that a large reservoir of molecular gas  dominates the dynamics of the region probed by our observations. This is consistent with the heavily dust-obscured UV emission observed in this galaxy. Similar to GN20, the physical properties of GN20.2a would point toward a major merger driven stage, likely approaching the final coalescence with substantial radio emission powered by an AGN, given the high gas surface density and extreme bright radio emission \citep{daddi09b,hodge13}. However, the complex optical morphology makes it difficult to understand the detailed physical process of star formation in this galaxy.  

Comparing the physical properties of GN20.2b with those of GN20 and GN20.2a, we find that GN20.2b exhibits comparable gas surface density and perhaps even distance with respect to the MS of the SFR-$M_\star$ plane with that observed in normal star-forming galaxy at $z\sim 2$. However, the optically thick UV emission and the high SFE observed in GN20.2b agree better with the case of a starburst. In combination with the analysis in Sect.~\ref{fraction}, the relatively old stellar age and short gas-consumption timescale of GN20.2b are likely to suggest that this galaxy has already passed its peak of star formation activity, probably observed at the decaying stage of a major merger. In addition, we note that the line width ($\sim 220 \pm 43$ km s$^{-1}$) of GN20.2b is much narrower (by a factor of $\sim$2.5--3.5) than those of GN20 and GN20.2a. Given the extreme obscuration of UV/optical emission in GN20.2a and GN20.2b, higher resolution and sensitivity CO imaging would be highly beneficial for the exploration of gas morphology and kinematics in these systems.

Despite the large uncertainties associated with the above analysis, our sample of three SMGs appears to be in different evolutionary stages: GN20 - a major merger undergoing final coalescence; GN20.2a - a major merger approaching final coalescence; and, GN20.2b - at the decaying stage of a merger-driven starburst. We consider that it could be reasonable to find starbursts at different stages in the same massive proto-cluster structure at $z=4.05$, as it would be hard to expect absolute synchronisation to much better than the 100~Myr timescale of a merger-driven starburst.

\section{Implications for the cosmic evolution of dust content in galaxies}~\label{implication}

The aim of this section is to study  the interplay between metals, gas, and dust grains, and the effect of metallicity evolution on observations of galaxies, by investigating the evolution of CO-luminosity-to-dust mass ratio and dust-to-stellar mass ratio across  cosmic time for starbursts as well as for MS galaxies, 
thus placing the observations of the GN20 proto-cluster SMGs in the general context.

\subsection{The evolution of $L'_{\rm CO}/M_{\rm dust}$}

\begin{figure}
\centering
\includegraphics[width=0.45\textwidth]{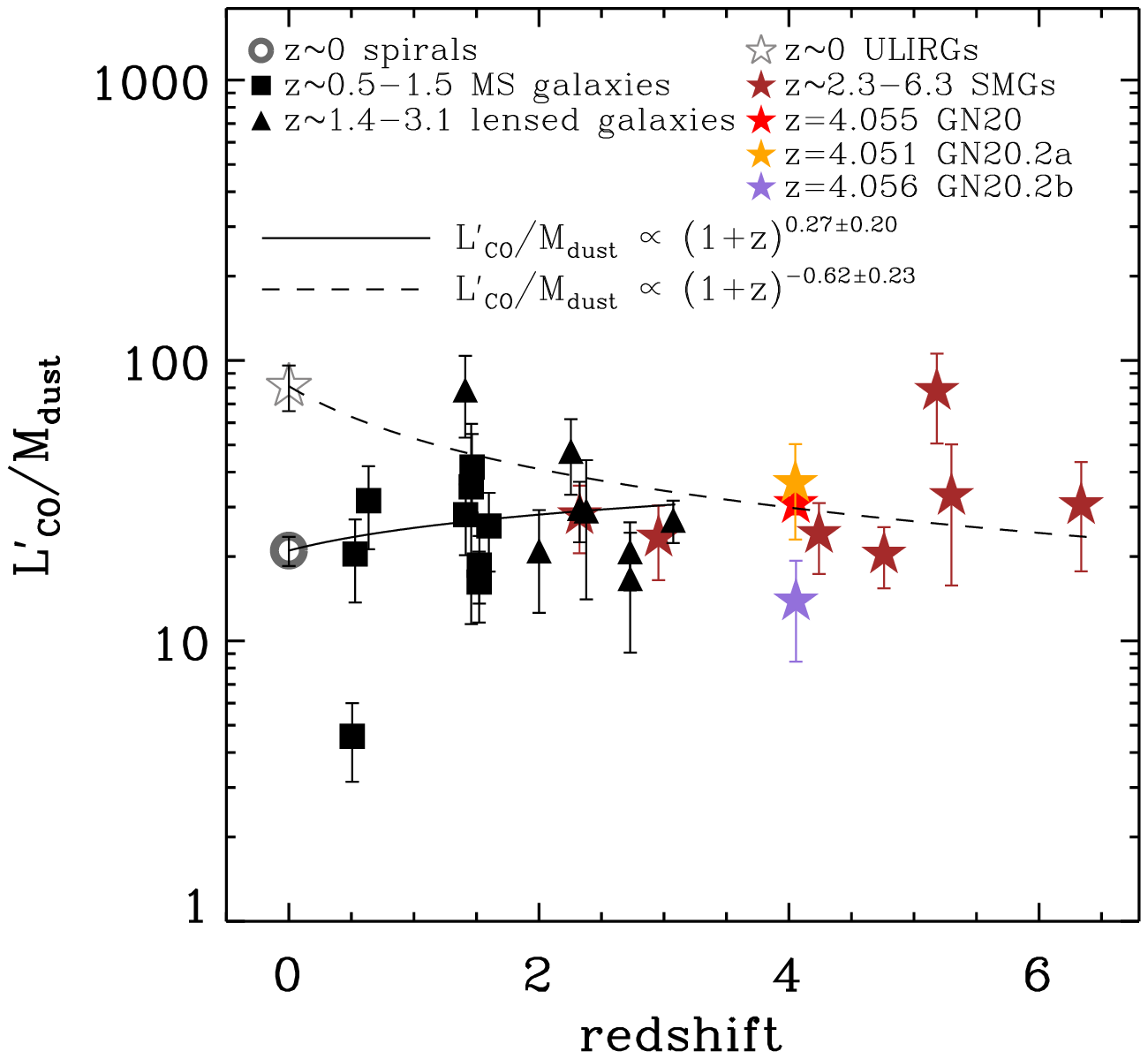}
\caption{Redshift evolution of $L'_{\rm CO}/M_{\rm dust}$ for MS galaxies and starbursts. At $z\sim 0$ we use the average value derived from the samples of \citet{leroy11} for the spirals (gray circle). Black squares represent individually detected normal star-forming galaxies at $z\sim 0.5-1.5$ \citep{daddi10,magdis12a}, while the black triangles denote lensed galaxies at $z\sim 1.4-3.1$ \citep{saintonge13}. The best fit to the normal galaxies yields a relation of {\bf $L'_{\rm CO}/M_{\rm dust} \propto (1+z)^{0.27\pm0.20}$} (solid line). In addition to GN20, GN20.2a, and GN20.2b in our sample at $z=4.05$, we also plot starbursts (grey and color stars) from the literature as follows: $z\sim 0$ ULIRGs \citep{downes98}; SMM J2135-0102 at $z=2.325$  \citep{swinbank10}; HSLW-01 at $z=2.957$ \citep{riechers11}; ID 141 at $z=4.243$ \citep{cox11}; LESS J033229.4-275619 at $z=4.760$ \citep{coppin10,gilli14}; HDF 850.1 at $z=5.183$ \citep{walter12}; AzTEC-3 at $z=5.298$ \citep{capak11,riechers10,dwek11}; HFLS3 at $z=6.34$ \citep{riechers13}. The best fit to the starburst galaxies gives a relation of {\bf $L'_{\rm CO}/M_{\rm dust} \propto (1+z)^{-0.62\pm0.23}$} (dashed line). \label{fig:co2dust}}
\end{figure}

We start by looking at the cosmic evolution of the dust mass to CO luminosity ratios, looking for empirical trends and trying to understand possible 
expectations for the relative evolution of these two quantities versus redshift, the effect of metallicity, and searching for possible different behaviours 
of MS versus SB galaxies.
Figure~\ref{fig:co2dust} shows the evolution of $L'_{\rm CO}/M_{\rm dust}$ with redshift for normal star-forming galaxies and starbursts. For MS galaxies, the sample available for our analysis consists of $z\sim 0$ local group galaxies from \citet{leroy11} and individually detected MS galaxies at $z\sim 0.5$ and $z\sim 1.5$ from \citet{magdis12a}. We also show $z\sim 1.4-3.1$ lensed star-forming galaxies from \citet{saintonge13}, although for lensed galaxies in general there is the possibility of a `lensing bias' affecting their selection (high surface brightness in the UV rest frame, and/or high intrinsic IR luminosity), so that it is not clear if they should be considered as typical MS galaxies, often times they could be starbursts (see also Tan et al. 2013). To reduce the possible uncertainty resulting from the assumption of excitation correction, for two of the UV-bright lensed sample galaxies, cB58 and Cosmic Eye,  we adopt the CO(1-0) luminosity measured by \citet{riechers10b} instead of the one extrapolated from CO(3-2) shown in \citet{saintonge13}. The dust masses of these galaxies are derived based on the DL07 model. For the starburst galaxies, besides our three $z=4.05$ SMGs, we add additional objects located at $z\sim 0-6.3$ from the literature \citep{downes98,coppin10,gilli14,cox11,swinbank10,riechers10,riechers11,riechers13,capak11,dwek11,walter12}. To be consistent with our targets, we also use DL07 models to estimate dust mass for the literature sources, except for ID 141 at $z=4.24$, LESS J033229.3-275619 at $z=4.76$, and HDF850.1 at $z=5.183$, for which we adopt the dust mass estimates in the literature after a correction factor of 2, given that the dust mass estimates based on a single-temperature modified blackbody model are on average a factor of $\sim$2 lower compared to those derived using DL07 model \citep{magdis12a,magnelli12a}. For the $z=6.3$ SMG \citep{riechers13}, our DL07 estimate of $M_{\rm dust}$ is much higher than the published one, 
and we adopt a log-averaged value which is within 0.2~dex from both estimates. 

The best fit to the MS galaxies yields a nearly flat ratio with redshift of  $L'_{\rm CO}/M_{\rm dust} \propto (1+z)^{0.27\pm0.20}$ from $z=0$ to 3.1, indicating that $L'_{\rm CO}/M_{\rm dust}$ remains roughly constant for normal galaxies. The best fit to the starburst galaxies gives a relation of {\bf $L'_{\rm CO}/M_{\rm dust} \propto (1+z)^{-0.62\pm0.23}$}. The GN20, GN20.2a and GN20.2b galaxies agree well with the general SB trend.
Local ULIRGs have a mean ratio of $L'_{\rm CO}/M_{\rm dust}$ higher than normal galaxies by factors of 3--4, but the  higher redshift starburst galaxies show consistent $L'_{\rm CO}/M_{\rm dust}$ with normal galaxies. These result do not change substantially, regardless of whether we include the lensed galaxies in the normal galaxy sample or in the starburst sample.

We now attempt a  simple interpretation of these trends, starting from the  MS galaxies.
We recall that both $L'_{\rm CO}$ and $M_{\rm dust}$ are intimately related to the gas mass, with $L'_{\rm CO}=M_{\rm H_2}/\alpha_{\rm CO}$ and $M_{\rm dust}=M_{\rm gas}/\delta_{\rm GDR}$, respectively. Therefore, the ratio of $L'_{\rm CO}/M_{\rm dust}$ can be expressed as $\delta_{\rm GDR}/\alpha_{\rm CO}$, provided that most of the gas contained in the region where the CO and IR luminosity arise is molecular. As discussed in Section~\ref{gdr}, H$_2$ is thought to dominate the gas mass of normal massive $M_\star >10^{10}M_\odot$ galaxies at low and high redshift, based on the large molecular gas fraction observed in high-$z$ MS galaxies \citep{daddi10,tacconi10}  and the theoretical arguments \citep[e.g.,][]{obreschkow09,bournaud11,lagos11}.
 It has been found that the gas-to-dust mass ratio is metallicity dependent with a relation of $\delta_{\rm GDR}\propto Z^{-1}$ for normal galaxies \citep[see][]{leroy11,sandstrom13}. Meanwhile, studies of CO-to-H$_2$ conversion factor have shown that $\alpha_{\rm CO}$ also  scales with metallicity on average as $\propto Z^{-1}$ for metallicities not much below $1/3$ solar \citep[see review by][]{bolatto13, genzel12, sargent13}.  Based on these derivations, the ratio of $L'_{\rm CO}/M_{\rm dust}$ should not really depend strongly on metallicity (hence neither on stellar mass nor on redshift), for massive galaxies. 

One might wonder if this expectation would change if the metallicity of massive galaxies in the high redshift Universe was substantially lower. 
Recent studies of dwarf galaxies have suggested  that both $\alpha_{\rm CO}$ and $\delta_{\rm GDR}$ could rise more rapidly at  metallicities $<10$~times smaller than solar \citep[e.g., with $\alpha_{\rm CO} \propto Z^{-2.4}$ in][]{schruba12}. Studies of gas-to-dust mass ratios of local galaxies over a large metallicity range reveal a broken power-law relation between $\delta_{\rm GDR}$ and metallicity \citep{remy14,fisher14}, with a steeper slope ($\propto Z^{-3}$) for dwarf galaxies with metallicities lower than $\sim 8.0$. Also in this case, the ratio of $L'_{\rm CO}/M_{\rm dust}$ would remain roughly constant as normal star forming galaxies with higher metallicity, i.e., the ratio of $L'_{\rm CO}/M_{\rm dust}$ changes little with metallicity and thus cosmic time. Of course, the ratio could instead decrease substantially in any regime in which the molecular to total hydrogen fraction was substantially smaller than unity.

Focusing now on the SB galaxies, 
 we already noted that the local ULIRGs show comparatively large $L'_{\rm CO}/M_{\rm dust}$ than that of normal galaxies  of similar metallicity \citep[see also Fig. 6 of][]{magdis12a}. It is interesting to try to interpret this empirical finding. As the ULIRGs are most likely H$_2$ dominated, this might be reconnected to either a larger $\delta_{\rm GDR}$ and/or a lower $\alpha_{\rm CO}$ than normal galaxies, at fixed  metallicities (i.e., stellar masses). In Fig. 5 of \citet{magdis12a}, the local ULIRGs are shown with on average lower $\alpha_{\rm CO}$ than the one derived based on the $\alpha_{\rm CO}$--$Z$ relation defined by normal galaxies, while the gas-to-dust mass ratios follow the local trend. Note that the metallicities derived in that study for ULIRGs are based on optical emission lines, and might be a biased estimate as they might be not very sensible to the most obscured regions.

As for the possible decline of the $L'_{\rm CO}/M_{\rm dust}$ ratio of SBs with redshifts, we suspect that it might be in part spurious. In fact, the local ULIRGs are more extreme starbursts than high-redshift ones, having  offsets from the MS of order of 10--30, while for example GN20 and GN20.2a have only six times higher SFR than MS galaxies of similar masses. Matched samples in terms of the distance to the MS would be required to investigate the possible differential redshift evolution in the properties of SB galaxies. 

All in all, these results suggest that the ratio of CO luminosity to dust mass remains approximately constant through cosmic time, albeit with larger ratio for starburst galaxies at $z=0$. A decline in metallicity could affect both quantities strongly, but in the same way at first approximation. Hence, similarly to the scenarios  briefly sketched in  \citet{tan13} for the evolution of the CO luminosity at high redshift (as for example measured respect to the average CO luminosity at fixed SFR in $z<2$ samples), in the case of rapidly declining metallicity in the distant Universe the dust mass of galaxies would be also substantially affected (and therefore their bolometric IR luminosities). In the next section we explore this issue in more detail, again trying to distinguish MS and SB galaxies. 

\subsection{The observed evolution of $M_{\rm dust}/M_{\star}$}

To guide the understanding of what could regulate the evolution of dust masses to high redshift, and to clarify respect to which physical properties of the galaxies the dust masses should be compared to assess their behaviour, it is useful to consider again the definition of $\delta_{\rm GDR}$, rewriting it in a more convenient form \citep[e.g.,][]{leroy11,magdis12a}:

\begin{equation}
M_{\rm dust}\sim\ 0.5\times Z \times\ M_{\rm gas}
\end{equation}
This uses a rough estimate of  50\% for the fraction of metals in dust, as on average 1\% of the gas is incorporated in dust in the local universe, while solar metallicity of $Z_\odot$=0.02 means that 2\% of the gas is in metals. A similar value of the dust-to-metal ratio ($\sim$0.5) was found by studies of extinction and metal column densities for a sample of $\gamma$-ray burst afterglows and quasar foreground absorbers at $z=0.1-6.3$ \citep{zafar13}. 
Dividing by stellar mass we obtain:
\begin{equation}
M_{\rm dust}/M_{\star}   \sim\ 0.5\times Z \times\ (M_{\rm gas}/M_{\star})
\label{eq:MDS}
\end{equation}
which shows that the dust to stellar mass ratio depends on the metallicity and the gas fraction in the galaxy, and is expected to be only weakly dependent on stellar mass at fixed redshift for MS galaxies (scaling like $M_{\star}^{-0.35}$ based on the dependencies of $Z$ and $M_{\rm gas}/M_{\star}$  discussed by Magdis et al. 2012a). Instead, at fixed stellar mass, the expected redshift evolution of $M_{\rm gas}/M_{\star}$ can be simply computed starting from the observed cosmic evolution of the sSFR in MS galaxies (see e.g., Fig.~19 in Sargent et al. 2013) and converting the SFR into gas masses using the Schmidt-Kennicutt law (S-K; SFR$\propto M^{1.2}_{\rm gas}$; see e.g., Daddi et al. 2010b; Genzel et  al. 2010; Sargent et al. 2013). The remainder of the redshift evolution 
of the $M_{\rm dust}/M_{\star}$ ratio is thus contained in the evolution of metallicity (or, at fixed stellar mass, of the mass-metallicity relation).

\begin{figure}[htbp]
\centering
\includegraphics[width=0.45\textwidth]{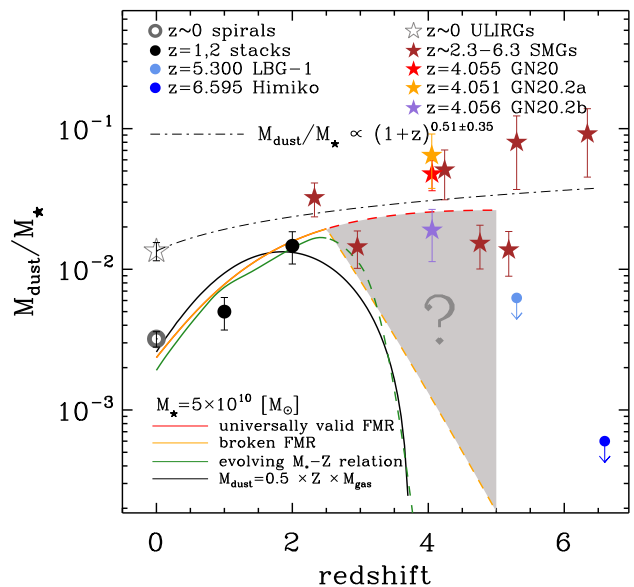}
\caption{Evolution of $M_{\rm dust}/M_\star$ as a function of redshift. The gray circle represents the averaged value of local normal galaxies from the sample of \citet{dacunha10a}, while black solid circles represent the stacking results at $z\sim$1 and $z\sim$2. The light blue and blue solid circles represent ``LBG-1'' and ``Himiko'', which are star-forming galaxies at $z$=5.300 and $z$=6.595, respectively. The black solid line shows the expected evolution of $M_{\rm dust}/M_\star$ with redshift based on the relation observed in the local universe, $M_{\rm dust}\sim 0.5\times M_{\rm gas}\times Z$. Predictions for the redshift-evolution of $M_{\rm dust}/M_\star$ of MS galaxies are shown for the case of an universally valid FMR/broken FMR/evolving $M_{\star}-Z$ relation at $z\geq$3 (red/orange/green lines; the $z>2.5$ extrapolations shown in colored dashed lines are observationally-unconstrained). The shaded region shows the range of $M_{\rm dust}/M_{\star}$ ratio beyond $z=2.5$ predicted by our models.  All measurements for normal galaxies and predictions have been normalized to a common mass scale of $M_\star = 5\times10^{10}\ M_\odot$. The star symbols indicate starbursts, same as in Fig.~\ref{fig:co2dust}. The black dot-dashed line is the best fit to the starbursts with a relation of $M_{\rm dust}/M_\star\ \propto (1+z)^{0.51\pm0.35}$.  \label{fig:dust2stellar}}
\end{figure}

In Fig.~\ref{fig:dust2stellar} we show measurements of $M_{\rm dust}/M_{\star}$ for MS and SB galaxies at different redshifts, normalized to a stellar mass of $5\times10^{10}M_\odot$ using the scaling discussed above. We show average values for $z\sim 0$ spiral galaxies \citep{dacunha10a} and stacked samples of $z\sim 1$ and $z\sim 2$ normal galaxies \citep{magdis12a}. The use of average samples is particularly useful as it should be representative of the typical  behaviour of MS 
galaxies, independently on  object-to-object fluctuations. Recent studies of nearby galaxies of the {\textit Herschel} Reference Survey derive a $M_{\rm dust}/M_{\star}$ of 3.5$\times$10$^{-3}$ at $z\sim 0$ \citep{cortese12,ciesla14}, in good agreement with our estimate for SDSS galaxies \citep{dacunha10a}. For typical MS galaxies, the dust-to-stellar mass ratio rises by a factor of few from $z=0$ to 2 \citep[consistent with the findings of][]{scoville14,santini14}. Similar trends have also been revealed for large samples of {\it Herschel}-ATLAS galaxies at $z<0.5$ \citep{dunne11}. Given that the lensed star-forming galaxies are intrinsically faint with low stellar mass ($M_\star \lesssim 10^{10}\ M_\odot$) and probably not representative of massive galaxies at those epochs, we did not compare the dust-to-stellar ratio of these galaxies with MS galaxies, although some lensed galaxies at $z\sim 1.5-3$ are shown to have similar $M_{\rm dust}/M_\star$ ratio to local star-forming galaxies \citep{sklias14}. For the SB galaxies we plot the same sample as in Fig.~\ref{fig:co2dust}  \citep{coppin10,gilli14,cox11,swinbank10,riechers10,riechers11,riechers13,capak11,dwek11,walter12}. The dust to stellar mass ratio of SB galaxies is found to be fairly flat with redshift, with the best fitting trend being  $M_{\rm dust}/M_{\star} \propto (1+z)^{0.51\pm0.35}$ (see Fig.~\ref{fig:dust2stellar}), indicative of still substantial metal enrichment at higher redshifts in these systems up to at least $z\sim 6$.  The dispersion of the residuals from the fit is larger than what expected from the measurement uncertainties, indicating intrinsic dispersion of at least a factor of two in the properties of the SB population. Again, the GN20 proto-cluster galaxies behave similarly to the rest of the SBs.

No dust mass measurements exist for indisputably normal, MS galaxies at $z>3$ (an attempt to study the evolution of $M_{\rm dust}/M_{\star}$ at  $3<z<4$ by stacking of normal MS galaxies will be presented in an upcoming paper by B\'{e}thermin et al. 2014, in prep.).
However, the recent ALMA observations of Himiko, a star-forming Ly-$\alpha$ selected galaxy at $z=6.595$ with SFR $\sim$ 100 $M_{\odot}$ yr$^{-1}$ and $M_{\star} \sim$ 1.5$\times$10$^{10}\ M_\odot$, reveal a significant deficit of dust content and [CII] 158$\mu$m emission \citep{ouchi13}, which is more than 30 times weaker than the one predicted by local correlations. With the 1.2 mm flux limit measured in \citet{ouchi13}, we estimate a dust mass assuming a radiation field $U$ to an appropriate value for a $z=6.6$ normal galaxy by extrapolating the relation of $\langle U \rangle \propto$ (1+$z$)$^{1.15}$ derived in \citet{magdis12a}.  This upper limit in dust mass, 
coupled with the substantial stellar mass of a few $10^{10}\ M_\odot$ in the system \citep{ouchi13} implies a $M_{\rm dust}/M_{\star}$ ratio that is more than an order of magnitude lower than the trend defined by the SB galaxies at $z>5$.  Analogously, we estimate a dust mass of a $z$=5.300 star-forming LBG called `LBG-1' based on the 1.0 mm flux limit observed with the ALMA \citep{riechers14b} and measure the stellar mass by fitting stellar population models to the rest-frame ultraviolet--optical photometry \citep{capak11}. We find an upper limit of $M_{\rm dust}/M_{\star}$ of 6.5$\times$10$^{-3}$ for this galaxy, significantly lower than the SB galaxies at the same epoch. These results are qualitatively similar to the recent observations toward two LBGs at $z>$3 reporting a deficiency in CO emission, supporting the scenario of a rapid decline in metallicity \citep{tan13}.

\subsection{A crucial ingredient: metallicity evolution in the distant Universe}

\begin{figure*}[htbp]
\centering
\scalebox{0.7}{\includegraphics*[68,23][640,295]{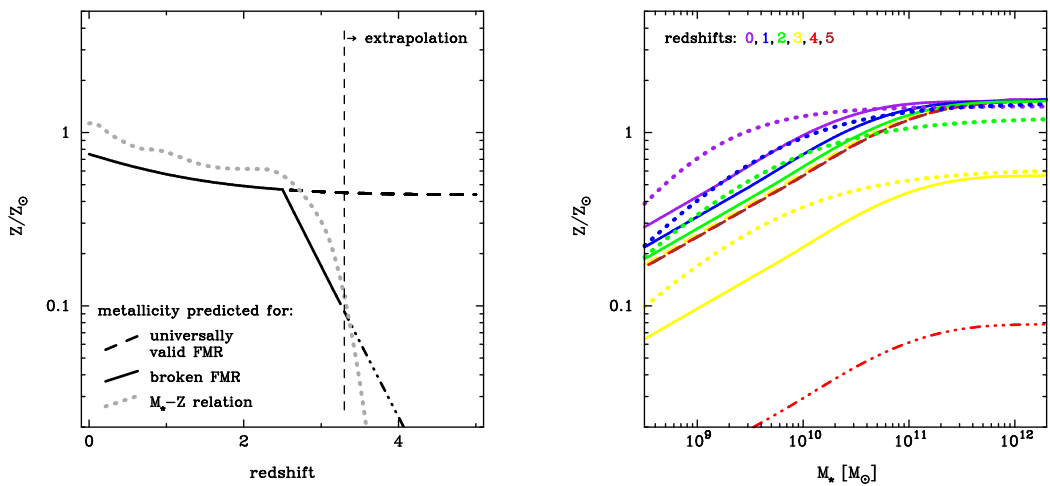}}
\caption{{\it Left}: the redshift evolution of the metallicity (normalized to solar) of a star-forming galaxy with $M_{\star}=5\times$10$^{10}$ $M_{\odot}$ for the case of an universally valid FMR/ broken FMR/ evolving $M_{\star}-Z$ relation at $z>2.5$ (dashed/solid/grey lines; the $z>3.3$ extrapolations are observationally unconstrained). {\it Right}:  the mass-metallicity relation between $z=0$ and 5 for the three cases  represented with same line styles as in the left panel. Lines are colour-coded according to the redshift. \label{fig:metallicity}}
\end{figure*}

We discuss in the following all these empirical findings. We start first with an attempt to interpret the behaviour of MS galaxies, and to predict their properties
at $z>3$ where observations are very scarce. Following the derivation of Eq.~\ref{eq:MDS} and the discussion above, together with the sSFR(z) evolution
as inferred e.g., in Sargent et al. (2013), the crucially needed ingredient is the 
evolution of metallicity for a given mass range (e.g., within 10 $<$log$M_{\star}/M_\odot<$ 11). We thus briefly review here the current understanding for the 
evolution in metallicity of massive galaxies through cosmic time. There are clearly two regimes that can be emphasized: above or below $z\sim 2.5-3$. 
From $z=0$ to 2.5 there is only a modest, factor of two or less decrease in the metallicity of massive galaxies \citep[e.g.,][many others]{erb06,tremonti04,zahid13}, see Fig.~\ref{fig:metallicity}-left. 
Recent studies on metal content of galaxies have also shown that galaxies up to $z\sim$2.5 follow the FMR defined locally with smooth variations \citep{mannucci10}.  The factor of a few increase of $M_{\rm dust}/M_{\star}$ for normal galaxies up to $z=2$ (Fig.~\ref{fig:dust2stellar}) is thus understood following Eq.~\ref{eq:MDS} in term of the rapid increase of gas fraction  \citep[e.g.,][]{daddi10,magdis12a,tacconi10,tacconi13}, which strongly overpowers for the metallicity decrement.

However, while
the exact evolution of average metallicity of massive galaxies at $z>3$ is still in debate and therefore highly uncertain, there are observational claims suggesting that things might change quite drastically at these high redshifts. Evidence for rapid metallicity evolution at $z>3$ has been proposed by studies of $z\sim 3.5$ LBGs \citep[e.g.,][]{maiolino08,mannucci10,sommariva12,troncoso13}, suggesting that the metallicity of massive $M_{\star}>10^{10}M_\odot$ galaxies might be decreasing fast, and the FMR no more valid over those redshift ranges (Fig.~\ref{fig:metallicity}-left). It is not clear if these measurements might be unrepresentative for 
typical $z>3$ massive galaxies, as they might be biased down in reddening (hence in metallicity) due to the UV-selection of the spectroscopic samples \citep[see][]{zahid13a}. Moreover,  these observational claims are in contrast with what is predicted by some theoretical  \citep[e.g.,][]{finlator08}  or heuristic \citep[e.g.,][]{lilly13} models.
Given the substantial uncertainties, for now we regard the range of scenarios as all plausible, and we explore in particular the implications within a reasonable range going from the case of stable metallicity of massive galaxies at $z>3$ (equivalent to non evolution in the FMR) up to the case in which 
the claim for rapid metallicity decrease \citep[e.g.,][]{troncoso13} is correct and representative for the whole population of $z\sim 3.5$ galaxies. 
The latter scenario can be computed either assuming a broken FMR framework, or parametrized in terms of the observed evolution of the mass-metallicity relation.
For the case of a broken FMR, we interpolate linearly with redshift the metallicity drop of 0.6 dex reported in \citet{mannucci10} between $z=2.5$ and $z=3.3$ and continue this trend to extrapolate beyond $z=3.3$. Similarly, we apply linear interpolation/extrapolation between $z=2.2$ and $z=3$ based on the $M_{\star}-Z$ relation of \citet{sommariva12}, \citet{troncoso13}, and \citet{zahid13} for the case of evolving $M_{\star}-Z$ relation.  In Figure~\ref{fig:metallicity}-right, we further plot the model predictions of the $M_{\star}$-$Z$ relation between $z=0$ and $z=5$ (color coded) for the above three different cases. For the scenarios involving the FMR, the SFR which was used to compute the metallicity at a given stellar mass is the SFR of the average MS galaxy. These plots illustrate how the mass-metallicity relation evolves with redshift under various hypothesis. For the case of an universally valid FMR, we find that the tracks at $z\geq 3$ are almost identical (Fig.~\ref{fig:metallicity}-right), with no indication of evolution of the $M_{\star}-Z$ relation above $z\sim$3, which is due to the fact that a plateau or slow rise is observed in the evolution of sSFR at $z>$3. For the case of accelerated metallicity evolution at $z>2.5$, the gas metallicity decreases rapidly with redshift. 

\subsection{Interpreting the evolution of $M_{\rm dust}/M_{\star}$}

Using again Eq.~\ref{eq:MDS}, we convert the different scenarios for the evolution of metallicity into related cosmic evolution of the  $M_{\rm dust}/M_{\star}$ ratio for normal star-forming galaxies with $M_{\star}=5\times$10$^{10}$ $M_{\odot}$  (Fig.~\ref{fig:dust2stellar}). The expected evolution trend of the $M_{\rm dust}/M_{\star}$ ratio derived directly from Eq.~\ref{eq:MDS} is shown in the black solid line. The metallicity is estimated based on the direct measurements of the mass-metallicity relation from optical surveys \citep{tremonti04,savaglio05,erb06,maiolino08} at $z<3.3$ and the extrapolation beyond higher redshift, while the ratio of $M_{\rm gas}/M_{\star}$ is computed from the observed cosmic evolution of the sSFR and of the integrated S-K relation \citep{sargent13}. 
The colored lines show the evolution trends predicted based on the above three scenarios for the evolution of metallicity at fixed stellar mass ($M_{\star}$=5$\times$10$^{10}$ $M_{\odot}$; see Fig.~\ref{fig:metallicity}), with dust mass converted from the gas mass computed similarly as above and assuming that the local $\delta_{\rm GDR}$--metallicity relation \citep{leroy11,magdis12a} is valid at high redshift. We find the predictions based on these three realisations are basically undistinguishable  up to $z\sim$2.5, with an increase evolution of $M_{\rm dust}/M_{\star}$ with redshift. For galaxies beyond $z\sim$2.5, however, the universally valid FMR would predict a mild increase of $M_{\rm dust}/M_{\star}$, while a rapid decline is expected for the case of accelerated evolution. 
The shaded region in Fig.~\ref{fig:dust2stellar} shows the predicted range of $M_{\rm dust}/M_{\star}$ ratio beyond $z$=2.5, emphasizing how uncertain the situation is. 

However, from Fig.~\ref{fig:dust2stellar} we find that the $M_{\rm dust}/M_{\star}$ ratios of LBG-1 and Himiko are significantly lower than the prediction from the universally valid FMR, consistent with the prediction assuming a rapid decline in metallicity at $z>$3. Unless there are something special with LBG-1 and Himiko, these two galaxies observations appear to support scenarios with declining metallicities for massive galaxies, at least in the early Universe.  

The scenarios with declining high-redshift metallicity would produce a broad 
peak of dust to stellar mass content in galaxies over at $z>1$ and up to 3 (or maybe higher, should a decline happen at earlier redshifts) that would allow one to term the $z\sim2$ range as the {\em epoch of dusty galaxies}, 
which would occur roughly at the same time as the peak in the SFRD. Recent studies of the IR to UV luminosity density ratio versus redshift depict a similar situation, with a broad $z\sim2$ peak characterising the ratio  \citep{cucciati12,burgarella13}. There seems to be a very similar behaviour in the dust-to-stellar mass ratio and the average dust attenuation in galaxies, at least to $z\sim2.5$, which could somewhat support again the scenarios with declining metallicities at high redshifts. 

We finally spend some words of caution noticing that, again following Eq.~\ref{eq:MDS}, a different evolution of gas-to-stellar mass ratio than what assumed here, could compensate in principle evolutionary trends in metallicity and influence the dust-to-stellar mass ratio evolution.  Current studies show no direct evidence of further increase of gas fraction at $z>$3 \citep[e.g.,][]{magdis12b,carilli13,sargent13,tan13}, and the sSFRs are revealed not to be rapidly rising as well \citep[e.g.,][]{bouwens12}. However it might well be that the gas-to-stellar mass ratio is actually rapidly rising at $z>3$ even if not seen by the sSFR \citep[e.g.,][]{dekel13}. The latter case would imply a decrease of star formation efficiency and deviation from the S-K law. 

We finish this discussion section with some consideration on the behaviour of the SB galaxies.  The local ULIRGs taken from the sample of \citet{dacunha10b} show substantially higher mean dust-to-stellar mass ratio than that of $z\sim 0$ disk galaxies, over a comparable stellar mass
range, suggesting that they have more dust or more metals (or probably both as dust grains form from the available metals in the ISM) than normal star forming galaxies at fixed stellar mass. 
The dust-to-stellar mass ratios derived for the SBs in our sample ranges between 0.01 and 0.09 with mean value of 0.04, in agreement with the $M_{\rm dust}/M_{\star}$ ratios of $z>1$ SMGs measured by \citet{rowlands14}. Starburst galaxies appear to saturate their $M_{\rm dust}/M_{\star}$ ratios at a value of few$\times 10^{-2}$, which  is unlikely to be a coincidence. This corresponds to the solar to few times solar metallicities of massive elliptical galaxies, into which these objects might be evolving once the effects of the undergoing merger event has faded.
If anything, the dust to stellar mass ratios for SBs might actually increase towards high redshifts, meaning that even in the early Universe SBs were substantially metal rich,  pointing towards the need of rapid metal enrichment in the starburst phase.
Studies of the origin of dust in galaxies using chemical evolution models found that, excepting asymptotic giant branch stars, supernovae and grain growth could be significant sources of dust to account for the high dust masses observed in star-bursting systems at high redshift \citep{rowlands14,michalowski10a,michalowski10b}. Direct evidence for the supernova to be a primary source of dust in the early Universe has been found by ALMA observations of Supernova 1987A, of which a remarkably large dust mass is revealed to be concentrated at the center of the remnant \citep{indebetouw14}. 

 All of these suggest that the measurement of $M_{\rm dust}/M_{\star}$ could  be a very powerful mean for distinguishing starbursts from normal galaxies at $z>4$, during the phase of early metal enrichment in starburst galaxies. 

\section{Summary}\label{summary}

We have presented  results from deep IRAM PdBI CO(4-3) and 1.2--2.2--3.3 mm continuum observations of the GN20 proto-cluster at $z=4.05$. The improved CO spectral profile of GN20, GN20.2a, and GN20.2b allow us to measure the line width more accurately and further constrain the dynamical mass. Combining with the ancillary multiwavelength photometry in the rest-frame UV, optical, and IR, we determine the stellar masses and dust properties of these three sub-mm galaxies (SMGs). With the measured stellar masses, dynamical masses, CO luminosities, dust masses, and indirect metallicity estimates, we inferred the value of conversion factor $\alpha_{\rm CO}$ in each of the three SMGs, using the dynamical modeling and the gas-to-dust ratio methods. Combining with literature data of normal galaxies and starbursts from local to high redshift, we discuss the effect of metallicity evolution on observations of dust and gas emission of galaxies across cosmic time. The main results and implications are summarized as follows:

1. All three SMGs are now  detected in CO(4-3) with high S/N ratios. The FWHM of the spectra derived from double Gaussian fits are 583$\pm$36 km s$^{-1}$, 760$\pm$180 km s$^{-1}$, and 220$\pm$43 km s$^{-1}$ for GN20, GN20.2a, and GN20.2b, respectively. With $L'_{\rm CO(4-3)}$ derived from our study and $L'_{\rm CO(2-1)}$ measured by \citet{carilli11}, we find CO(4-3)/CO(2-1) line ratio of $\sim$0.4 for these three SMGs, which is consistent with the mean ratio ($\sim$0.48$\pm$0.10) measured for SMGs at $z\sim2$--4 \citep{bothwell13}. 

2. We report  3.3 mm continuum detections in GN20.2a and GN20.2b for the first time, and use continuum measurements at 1.2 and 2.2 mm (Carilli et al. 2010; Riechers et al. 2014, in prep.). The dust masses measured from the far-IR SED fitting are $2.1-5.2\times10^9$ M$_\odot$. The IR-to-radio luminosity ratios for GN20 ($q$=2.41$\pm$0.05) and GN20.2b ($q$=2.60$\pm$0.13) are found to be comparable to the local value ($\sim$2.6), while GN20.2a shows a relatively low value ($q$=1.72$\pm$0.04), suggesting that this galaxy is radio-excess with large amounts of radio emission likely powered by an AGN (see also Daddi et al. 2009).

3. We find that the value of $\alpha_{\rm CO}$ inferred from gas-to-dust ratio method is consistent with the one derived based on the dynamical mass for each galaxy. The $\alpha_{\rm CO}$ derived for these three SMGs ($\sim$1.3-2.8 $M_\odot$ (K km s$^{-1}$ pc$^2$)$^{-1}$) are found to be consistent
with  the typical value determined for local ULIRGs, but might be well below the value appropriate for normal galaxies at similar redshifts. The high gas fraction ($\sim$40\%-80\%) of these three SMGs at $z$=4.05 are found to be comparable to SMGs at $z$=2--4 and high redshift normal galaxies.

4. Our study clearly distinguishes GN20, GN20.2a  (and likely GN20.2b) as starbursts from normal star-forming galaxies by comparing observed physical properties between these galaxies. For GN20 and GN20.2a, the large sSFR-excess (sSFR/sSFR$_{\rm MS} \sim 6$) compared to the normal galaxies at similar epochs place these SMGs as outliers above the main sequence.  The extremely large value of SFR$_{\rm IR}$/SFR$_{\rm UV}$ for GN20 and GN20.2a are consistent with the large offset between CO positions and optical counterparts (see Fig.~\ref{counterparts}), suggesting that the UV/optical emission is heavily obscured beyond optically thick dust. Although GN20.2b situates within the MS scatter with sSFR/sSFR$_{\rm MS}\sim 2.4$, both the large value of SFR$_{\rm IR}$/SFR$_{\rm UV}$ ($\sim 6$) and short gas-consumption time-scales ($\sim$116 Myr)  suggest that this galaxy could also be a starburst. 

5. We find that these three SMGs are likely to experience different evolutionary stages of starburst activity. Compared with GN20 and GN20.2a, GN20.2b displays relatively smaller sSFR-excess, older stellar age, and lower CO excitation. All these properties suggest that GN20.2b is probably observed at a decaying stage of a major merger, while GN20 and GN20.2a are likely to undergo and approach the final coalescence with intense starburst, respectively. 

6. We compile a variety of archival data of normal galaxies and starbursts in order to investigate the effect of metallicity evolution on observations of galaxies across the cosmic time. For normal galaxies, the ratio of $L'_{\rm CO}/M_{\rm dust}$ is found to be almost constant from $z=0$ to 3.1. And the same appear to be the case for high-z starbursts, implying that both the CO and dust emission could be affected in the same way by the metals in the ISM. 

7. We calculate simple models for the dust emission of normal galaxies based on their properties and find a rapid decrease of dust emission for $z>3$ normal galaxies at a given stellar mass if metallicities indeed decrease rapidly for these galaxies. The model predictions are well-matched to the measurements at $z<2.5$, with a trend of increasing $M_{\rm dust}/M_{\star}$ with redshift. While no dust detection is available for indisputably normal galaxies beyond $z=2.5$, the model based on the assumption of fast metallicity evolution predicts a sharp decline of dust emission for normal galaxies at $z>2.5$. In contrast, the starburst galaxies show an increase of $M_{\rm dust}/M_{\star}$ at high redshift, providing evidence for rapid early metal enrichment in these systems. The different behaviours of normal galaxies lead to a significantly lower $M_{\rm dust}/M_{\star}$ compared to starbursts at $z>4$, implying that the comparison of $M_{\rm dust}/M_{\star}$ ratio could also be used as a powerful tool for distinguishing starbursts from normal galaxies at $z>4$.

\begin{acknowledgements}
This work was based on observations carried out with the IRAM PdBI, supported by INSU/CNRS(France), MPG(Germany), and IGN(Spain). The authors  thank Rapha\"{e}l Gobat for assistance with the  SED fitting of LBG-1, and we thank the anonymous referee for useful comments and suggestions which greatly improved this manuscript. Q.T., E.D., and M.T.S. acknowledge the support of the ERC- StG UPGAL 240039 and ANR-08-JCJC-0008 grants. Q.T. was supported by CAS and CNRS, and partially supported by the NSFC grant \#11173059, \#11390373, and the CAS grant \#XDB09000000.

\end{acknowledgements}





\bibliographystyle{aa.bst}
\bibliography{reference}

\clearpage







\end{document}